\documentclass[12pt]{article}
\usepackage{amssymb,graphicx,amsmath,array,verbatim,amscd}%,cite}
\usepackage[usenames]{color}

\makeatletter
\@addtoreset{equation}{section}

\renewcommand{\thefootnote}{\fnsymbol{footnote}}
\makeatother

\newcommand {\beq}{\begin{eqnarray}}
\newcommand {\eeq}{\end{eqnarray}}
\def\p{\partial}
\newcommand{\NF}{N_{\rm F}}

\newcommand{\vs}[1]{\vspace{#1 mm}}
\newcommand{\hs}[1]{\hspace{#1 mm}}
\newcommand{\bpm}{\begin{pmatrix}}
\newcommand{\epm}{\end{pmatrix}}

\newcommand{\C}{\mathbb{C}}

\newcommand{\tr}{{\rm Tr}}
\newcommand{\D}{\mathcal D}

\newcommand{\ba}{\left( \begin{array}}
\newcommand{\ea}{\end{array} \right)}
\newcommand{\be}{\begin{equation}}
\newcommand{\ee}{\end{equation}}
\newcommand{\bea}{\begin{eqnarray}}
\newcommand{\eea}{\end{eqnarray}}
\newcommand{\beann}{\begin{eqnarray*}}
\newcommand{\eeann}{\end{eqnarray*}}

\newcommand{\Z}{\mathbb{Z}}

\begin{document}
%\maketitle
\thispagestyle{empty}
\begin{flushright}
IFUP-TH/2012-05 \\
KUNS-2393 \\
YGHP-12-46 \\
April, 2012
\end{flushright}
\vspace{1mm}
\begin{center}
{\LARGE \bf Higher Derivative Corrections \vs{3} \\
to Non-Abelian Vortex Effective Theory} \\

\vspace{0.5cm}
{\normalsize\bfseries
Minoru Eto${}^{1}$, Toshiaki Fujimori${}^{2}$, 
Muneto Nitta${}^{3}$,\\ Keisuke Ohashi${}^{4}$ and
Norisuke Sakai${}^{5}$}
\footnotetext{
Email addresses: \tt
meto(at)riken.jp, 
toshiaki.fujimori(at)pi.infn.it, \\
nitta(at)phys-h.keio.ac.jp, 
ohashi(at)gauge.scphys.kyoto-u.ac.jp,
sakai(at)lab.twcu.ac.jp}

\vskip 1.5em
{\it\small 
$^1$ Department of Physics, Yamagata University, Yamagata 990-8560, Japan}\\
{\it\small 
$^2$ INFN, Sezione di Pisa, 
Largo B.~Pontecorvo, 3,  Ed.~C, 56127 Pisa, Italy, and}\\
{\it\small 
Department of Physics,``E. Fermi'', University of Pisa, 
Largo B.\,Pontecorvo,\,3, Ed.\,C, 56127 Pisa, Italy}
\\{\it\small
$^3$ Department of Physics, and Research and Education 
Center for Natural Sciences,}\\  
{\it\small 
 Keio University, Hiyoshi 4-1-1, Yokohama, Kanagawa 223-8521, Japan}\\
{\it\small 
$^4$ Department of Physics, Kyoto University, Kyoto 
606-8502, Japan}\\
{\it\small 
$^5$ 
Department of Mathematics, Tokyo Woman's Christian University, 
%Zenpukuji, Suginami, 
Tokyo 167-8585, Japan
}\\
\vspace{10mm}

%%%%%%%%%%%%%%%%%%%%%%%%%%%%%%%%%%%%%%%%%%%%%%%%%%%%%%%%%%%%%%%%%%%%%%%%%%%
\begin{abstract}
We give a systematic method to calculate higher derivative 
corrections to low-energy effective theories of solitons, 
which are in general nonlinear sigma models on the moduli 
spaces of the solitons. 
By applying it to the effective theory of 
a single BPS non-Abelian vortex 
in $U(N)$ gauge theory with $N$ fundamental Higgs fields, 
we obtain four derivative corrections to 
the effective sigma model on the moduli 
space ${\mathbb C} \times {\mathbb C}P^{N-1}$. 
We compare them with the Nambu-Goto action 
and the Faddeev-Skyrme model. 
We also show that Yang-Mills instantons/monopoles trapped 
inside a non-Abelian vortex membrane/string 
are not modified in the presence of higher derivative terms.
\end{abstract}

\end{center}
\vfill
\newpage
\tableofcontents
\thispagestyle{empty}
\newpage

\setcounter{footnote}{0}
\renewcommand{\thefootnote}{\arabic{footnote}}
\setcounter{page}{1}
%%%%%%%%%%%%%%%%%%%%%%%%%%%%%%%%%%%%%%%%%%%%%%%%%%%%%%%%%%%%%%%%%%%%%%%%%%%
\section{Introduction} 

Solitons are smooth localized solutions of 
nonlinear partial differential equations, 
and they are ubiquitous in diverse fields in physics and mathematics. 
Topological solitons are smooth localized solutions 
of field equations in quantum field theories, 
and their stability is ensured by topological charges. 
A particularly interesting class of topological solitons 
is Bogomol'nyi-Prasad-Sommerfield (BPS) solitons \cite{Bogomolny:1975de}, 
which saturate minimum energy bounds called the Bogomol'nyi bounds. 
They are the most stable solutions with given topological 
numbers and satisfy first order partial differential 
equations called BPS equations.  
Prominent examples of such BPS solitons are 
instantons in Euclidean Yang-Mills theory \cite{Belavin:1975fg}, 
`t Hooft-Polyakov monopoles \cite{'tHooft:1974qc} 
in the Bogomol'nyi limit \cite{Bogomolny:1975de},
and Abrikosov-Nielsen-Olesen (ANO) vortices \cite{Abrikosov:1956sx} 
at the critical coupling (a standard reference is \cite{Manton:2004tk}).  
Recent examples of BPS topological solitons in gauge theories are 
BPS kinks (domain walls) in 
Abelian \cite{Abraham:1992vb} and non-Abelian \cite{Isozumi:2004jc}
gauge theories.

While they preserve half of supersymmetry and are quantum 
mechanically stable in supersymmetric gauge theories 
\cite{Witten:1978mh} on one hand,  
one of the most important features of BPS solitons is that 
there exist no static forces among them 
and consequently a continuous family of configurations 
with degenerate energy is allowed. 
As a result, generic solutions of BPS solitons 
contain collective coordinates, 
called moduli parameters as integration constants.
The moduli parameters of BPS solitons 
parametrize the space of whole solutions of BPS equations, 
called the moduli space. 
Although there is no static force among BPS solitons, 
they non-trivially scatter each other when they are moving. 
While such dynamics was studied by computer simulations, 
it is very difficult to study it by means of a fully analytic approach. 
The seminal idea of Manton \cite{Manton:1981mp} is that 
when solitons move slowly, their dynamics can be described 
as geodesics on their moduli space. 
This low-energy approximation is now called the moduli, geodesic, 
or Manton approximation.
After his work, the moduli spaces are recognized as 
the most important objects associated with BPS solitons \cite{Manton:2004tk}. 
The Atiyah-Hitchin metric on two BPS monopoles 
is one of the most prominent examples \cite{Atiyah:1985dv}. 
By examining geodesics on it, 
one can study the scattering of two BPS monopoles 
and find some interesting phenomena such as 
the right angle scattering in head on collisions. 
The moduli space of BPS monopoles was 
further studied in \cite{Gibbons:1995yw,Lee:1996kz}.
The moduli space and soliton scattering were studied 
for ANO vortices in the BPS limit 
\cite{Taubes:1979tm,Ruback:1988ba,Shellard:1988zx,
Samols:1991ne,Manton:2002wb,Chen:2004xu,Krusch:2009tn} 
and for BPS domain walls \cite{Tong:2002hi}.
More recently, the moduli space dynamics has been 
successfully applied even to 1/4 BPS composite solitons 
such as domain wall networks \cite{Eto:2006bb} 
and strings stretched between parallel domain walls \cite{Eto:2008mf}. 
Although there has been a lot of remarkable progress 
in study of the moduli space of BPS solitons, 
the analysis of their dynamics has been restricted 
to the leading order of the low-energy (small-velocity) limit, 
except for a few examples \cite{Eto:2005cc,Hata:2010vj}.

Non-Abelian vortices were found recently in ${\cal N}=2$ 
supersymmetric $U(N)$ QCD \cite{Hanany:2003hp,Auzzi:2003fs}. 
Although there has been a lot of progress 
in the study of non-Abelian vortices \cite{review,Eto:2006pg}, 
here we concentrate on their moduli space.
A single non-Abelian vortex has moduli space
${\mathbb C} \times {\mathbb C}P^{N-1}$ where 
the former denotes the position and the latter, 
called orientational moduli, comes from the $SU(N)$ vacuum 
symmetry broken in the presence of the vortex 
\cite{Hanany:2003hp,Auzzi:2003fs}. 
One of the most important facts about the orientational moduli 
is that confined monopoles and trapped Yang-Mills instantons 
in the Higgs phase are realized, respectively, 
as kinks \cite{Tong:2003pz} and 
lumps \cite{Eto:2004rz,Fujimori:2008ee} 
in the vortex world-sheet ${\mathbb C}P^{N-1}$ sigma model. 
This world-sheet effective theory provides 
a physical explanation for the relationship between 
BPS spectra in two-dimensional $\mathcal N = (2,2)$ sigma models 
and four-dimensional $\mathcal N = 2$ supersymmetric QCD 
\cite{Dorey:1998yh,Shifman:2004dr}.
The moduli space of multiple vortices with full moduli 
parameters was completely determined (without metric) 
in \cite{Eto:2005yh,Eto:2006pg,Eto:2006cx};  
The moduli space for $k$ separated vortices is a 
$k$-symmetric product 
$({\mathbb C} \times {\mathbb C}P^{N-1})^k/{\mathcal S}_k$ 
of the single vortex moduli space 
 \cite{Eto:2005yh} while the whole space is regular.
General formula for the moduli space metric and its K\"ahler 
potential were given in \cite{Eto:2006uw}.
The metric of the moduli space on the coincident vortices
was found \cite{Eto:2006db,Eto:2010aj} and was applied to 
low-energy dynamics of two non-Abelian vortices in head-on collision, 
reconnections of two non-Abelian cosmic strings \cite{Eto:2006db} 
and flux matching of vortex-monopole composite \cite{Eto:2006dx}.
Recently the moduli space metric of 
multiple non-Abelian vortices has been finally obtained 
for well separated vortices \cite{Fujimori:2010fk}, 
and their low-energy dynamics has been examined \cite{Eto:2011pj}. 

The purpose of this paper is to propose a systematic method 
to study dynamics of BPS solitons moving with higher 
velocities beyond the Manton approximation; 
we give a general scheme to calculate derivative corrections 
and apply it to obtain four-derivative corrections 
to the low-energy effective theory on a single non-Abelian vortex.
In the spirit of the low-energy effective action, 
the leading order terms can be obtained 
by integrating out massive modes. 
The lowest order terms take the form of nonlinear sigma models 
with two derivative terms for light or massless fields, 
typically Nambu-Goldstone modes. 
A famous example is the chiral Lagrangian or 
chiral perturbation theory for pions 
which are the Nambu-Goldstone bosons of chiral symmetry breaking. 
The next-leading terms consist of four or higher derivative corrections, 
which are the focus of our study.  
In order to obtain four derivative terms, 
we solve equations for massive modes and eliminate them order by order. 
As a concrete example, we consider a single non-Abelian vortex. 
We obtain four-derivative corrections 
for the translational moduli ${\mathbb C}$, 
the orientational moduli (the ${\mathbb C}P^{N-1}$ model) 
and their mixing terms.

%%%%%%%%%%%%%%%%%
In the literature a different expansion is known: 
an expansion from the Nambu-Goto action \cite{Nambu:1974zg} 
in powers of (the inverse of) the width of strings. 
The effective action of a string in the thin limit 
can be described by the Nambu-Goto action \cite{Forster:1974ga}.
The finite-width correction in general takes 
the form of extrinsic curvature squared, 
which is called stiffness or rigidity of string 
\cite{Polyakov:1986cs}. 
Dynamics of a string with the correction term 
was studied \cite{Curtright:1986vg}.
The explicit calculation of the rigidity term 
was performed in the Abelian Higgs model by many authors   
in the context of cosmic strings and QCD strings \cite{higher}. 
This expansion is in powers of $\# (\partial) - \# (X)$, 
where we denote the number of derivatives and fields as 
$\# (\partial)$ and $\# (X)$, respectively. 
The leading term, the Nambu-Goto action, 
is the zeroth order term in the series, 
and thus contains the same number of derivatives and fields.
The next-leading term, the stiffness term, 
contains two more derivatives than the number of fields 
starting from ${\cal O}(\partial^6 X^4)$. 
One advantage of this width expansion is that even the leading term,
the Nambu-Goto action, contains infinite number of derivatives 
so that it can describe large fluctuations or bends of a string.
However the expansion in powers of $\# (\partial) - \# (X)$ is 
not applicable for dynamics of moduli in general; 
for instance multiple solitons or 
even a single non-Abelian soliton with internal degrees of freedom. 
--- On the other hand, our expansion is purely a derivative expansion 
commonly used in the literature of field theories. 
It is valid at low energy and even the leading term,  
the nonlinear sigma model, can describe dynamics of multiple solitons.

In the present paper, we will discuss 
the higher derivative corrections to 
the low-energy effective theory on a non-Abelian vortex world-volume. 
By a symmetry argument, one can easily write down 
the generic form of the higher derivative terms 
and find that the corrections cause instability for the lumps 
(sigma model instantons) which are identified with 
the Yang-Mills instantons trapped inside a vortex. 
According to the Derrick's scaling argument, 
the size of a lump expands 
in the presence of the generic higher derivative terms.
Since the lumps are responsible 
for non-perturbative effects 
in the vortex world-sheet effective theory, 
one may think that the correspondence between 
the BPS spectra in 2d and 4d \cite{Dorey:1998yh,Shifman:2004dr} 
would be modified by the higher derivative corrections. 
However, as we will see in Sec.\,\ref{sec:inst}, 
the lump solutions are still stable 
if the higher derivative terms have a specific form. 
In this paper, we will calculate the explicit form of 
the higher derivative terms 
and find that they do not modify the lump solutions. 

This paper is organized as follows. 
We first review non-Abelian vortices and 
their moduli space in Sec.\,\ref{sec:NAvor}.
After illustrating our method for the derivative expansion 
in a simple example of classical mechanics in Sec.\,\ref{sec:prelim},
we develop a systematic method to obtain derivative corrections 
to the effective action for non-Abelian vortices in Sec.\,\ref{sec:deriv}.
In Sec.\,\ref{sec:com} we compare our result 
for the effective action of a single vortex with previously known models, 
the Nambu-Goto action and the Faddeev-Skyrme model \cite{Faddeev:1996zj}. 
In Sec.\,\ref{sec:inst} we discuss that 
the four derivative terms do not modify the Yang-Mills 
instanton solutions trapped inside a non-Abelian vortex 
\cite{Eto:2004rz,Fujimori:2008ee}. 
In Sec.\,\ref{sec:mass}, we discuss higher order corrections 
to the vortex effective action in a mass deformed model 
and show that the classical BPS spectrum is not modified by 
the higher order corrections.
Sec.\,\ref{sec:sum} is devoted to summary and discussion.

\newpage

%%%%%%%%%%%%%%%%%%%%%%%%%%%%%%%%%%%%%%%%%%%%%%%%%%%%%%%%%%%%%%%%%%%%%%%%%%%
\section{The non-Abelian vortices}  \label{sec:NAvor}
In this section, we briefly review the non-Abelian vortices 
and summarize the basic tools to describe 
the moduli space of BPS configurations. 

Let us consider the $U(N)$ gauge theory 
in $(d+1)$-dimensional spacetime 
with gauge field $W_\mu$
and $\NF=N$ Higgs fields $H$ ($N$-by-$N$ matrix)
in the fundamental representation. 
The Lagrangian of our model takes the form 
\beq
\mathcal L &=& \tr \left [ - \frac{1}{2g^2} F_{\mu \nu} F^{\mu \nu}
+ \D_\mu H (\D^\mu H)^\dagger - \frac{g^2}{4} (H H^\dagger - v^2 \mathbf 1_N)^2 \right], \label{eq:action} 
\eeq
where $g$ is the gauge coupling constant and 
$v^2$ is the Fayet-Iliopoulos parameter. 
We use the almost minus metric $\eta_{\mu\nu} = (+1,-1,-1,\cdots,-1)$. 
Although we can choose different gauge coupling constants 
for the $U(1)$ and $SU(N)$ parts of the gauge group, 
we set them equal for notational simplicity. 
Our notation for the covariant derivative and the field strength is
\beq
\D_\mu H = (\p_\mu + i W_\mu) H, \hs{10} 
F_{\mu \nu} = \p_\mu W_\nu - \p_\nu W_\mu + i [ W_\mu , W_\nu ]. 
\eeq
As is well known, 
the Lagrangian can be embedded into 
a supersymmetric theory with eight supercharges.
The vacuum condition is solved by
\beq
H = v \, \mathbf 1_N.
\label{eq:H_VEV}
\eeq
This vacuum expectation value (VEV) of the Higgs field 
completely breaks the gauge symmetry, 
whereas the following $SU(N)_{\rm C + \rm F}$ diagonal symmetry 
remains unbroken
\beq
H ~\rightarrow~ U_{\rm C} H U_{\rm F}, \hs{10} U_{\rm C}^\dagger = U_{\rm F} \in SU(N)_{\rm C +F}.
\eeq
Let us consider the non-Abelian vortices 
in this color-flavor locked Higgs vacuum. 
We assume that the vortices are localized in the $z = x_{d-1} + i x_d$ plane. 
The topological sectors of the field configurations are 
classified by the vorticity, i.e. 
the magnetic flux of the overall $U(1)$ gauge group
\beq
k ~\equiv~ - \frac{1}{2\pi} \int dz \wedge d \bar z \ \tr \, F_{z \bar z}, \hs{10} k \in \Z.
\label{eq:vorticity}
\eeq
For a given vorticity $k$, 
the tension (the energy per unit volume) 
of static configurations is bounded from below as
\beq
T \hs{-2} &=& \hs{-2} \int d^2 x \, \tr \Bigg[ 4 \D_{\bar z} H (\D_{\bar z} H)^\dagger + \frac{4}{g^2} \left| i F_{z \bar z} - \frac{g^2}{4} ( v^2 \mathbf 1_N - H H^\dagger ) \right|^2 - 4 \D_{[z} ( \D_{\bar z]} H H^\dagger ) + 2 v^2 i F_{z \bar z} \Bigg] \notag \\
&\ge& 2 \pi v^2 k,
\label{eq:BPS_bound}
\eeq
where we have assumed that $\D_i H \rightarrow 0$ 
at the infinity $|z| \rightarrow \infty$
so that the third term does not contribute to the energy. 
This Bogomol'nyi bound is saturated if the following equations are satisfied
\beq
\D_{\bar z} H = 0, \hs{10}
i F_{z \bar z} = \frac{g^2}{4} ( v^2 \mathbf 1_N - H H^\dagger). 
\label{eq:BPS1} 
\eeq
These equations are the BPS equations for non-Abelian vortices 
and describe the configurations of static BPS vortices 
which are minimal energy configurations in a given topological sector.
Since the same equations can be derived by imposing the condition 
that one half of supersymmetry is preserved, 
Eq.\,(\ref{eq:BPS1}) is also called 1/2 BPS equations. 
We can show that any of the BPS configurations is 
a solution of the following full equations of motion of the system
\beq
0 &=& \D_\mu \D^\mu H + \frac{g^2}{2} ( H H^\dagger - v^2 \mathbf 1_N ) H, 
\label{eq:eom_full1} \\
0 &=& \frac{2}{g^2} \D^\mu F_{\mu \nu} + i \left[ H ( \D_\nu H )^\dagger - (\D_\nu H) H^\dagger \right]. \label{eq:eom_full2}
\eeq
In order to describe the moduli space of the BPS configurations, 
it is convenient to write the BPS solution as
\beq
H = v \, S^{-1} H_0, \hs{10} 
W_{\bar z} = - i S^{-1} \bar \p S,
\label{eq:BPSsol}
\eeq
where $H_0$ and $S$ are $N$-by-$N$ matrices. 
Then, the first BPS equation in Eq.\,(\ref{eq:BPS1}) becomes
\beq
\p_{\bar z} H_0 = 0.
\eeq
Therefore all the entries of the $N$-by-$N$ matrix $H_0$ 
are arbitrary holomorphic polynomials. 
The matrix function $S \in GL(N,\C)$ 
is determined from the second equation in Eq.\,(\ref{eq:BPS1}), 
which can be rewritten into the following equation for 
$\Omega \equiv SS^\dagger$ : \cite{Isozumi:2004vg,Eto:2005yh}
\beq
\frac{4}{g^2 v^2} \p_{\bar z} ( \Omega \p_z \Omega^{-1} ) 
= H_0 H_0^\dagger \Omega^{-1} - \mathbf 1_{N}.
\label{eq:master}
\eeq
For a given $H_0$, this equation can be viewed as
a non-linear differential equation for $\Omega$. 
The boundary condition for $\Omega$ should be chosen so that 
the energy density vanishes at the spatial infinity $|z| \rightarrow \infty$. 
In general, this condition is solved by
\beq
\Omega ~\underset{|z| \rightarrow \infty}{\longrightarrow}~ H_0 H_0^\dagger. 
\label{eq:boundary_Omega}
\eeq
The matrix $S$ can be determined from the solution $\Omega$ 
uniquely up to the unphysical $U(N)$ gauge transformation 
$S \rightarrow S u(x)$ (with $u(x) \in U(N)$). 
Solving Eq.\,\eqref{eq:master} for an arbitrarily chosen $H_0(z)$, 
we can obtain a BPS vortex solutions through Eq.\,\eqref{eq:BPSsol}. 
In other words, the matrix $H_0(z)$ classifies all the BPS 
configurations. 
Hence the holomorphic matrix $H_0(z)$ is called ``the moduli matrix" 
and Eq.\,\eqref{eq:master} is called ``the master equation" for vortices. 

The matrices $H_0(z)$ and $\Omega$ are invariant under 
the original $U(N)$ gauge transformation. 
On the other hand, the original fields $H$ and $W_{\bar z}$ are 
invariant under ``the $V$-transformations" defined by
\beq
H_0(z) \rightarrow V(z) H_0(z), \hs{7} 
\Omega \rightarrow V(z) \, \Omega \, V(z)^\dagger, \hs{10} 
V(z) \in GL(N,\C).\label{eq:V-trf}
\eeq
This is the gauge symmetry of the master equation Eq.\,\eqref{eq:master} 
which does not change the physical quantities. 
Therefore, the solutions of the BPS equations are completely classified by 
the equivalence classes of the moduli matrix 
\beq
H_0(z) ~\sim~ V(z) H_0(z).
\label{eq:V-transf}
\eeq
This implies that the parameters contained 
in an appropriately gauge fixed moduli matrix $H_0$ can be 
interpreted as the moduli parameters of the BPS configurations. 

Let us next discuss how the moduli matrix $H_0$ determines 
the topological sector of the corresponding solution. 
From Eq.\,\eqref{eq:BPSsol}, we can see that 
the magnetic flux of the overall $U(1)$ is given in terms 
of $\Omega$ by
\beq
\tr \, F_{z \bar z} = - i \p_{\bar z} \p_z \log \det \Omega.
\eeq
By using the Stokes' theorem and 
the boundary condition Eq.\,\eqref{eq:boundary_Omega}, 
we find that the vorticity Eq.\,\eqref{eq:vorticity} is given by
\beq
k ~=~ \frac{1}{4 \pi i} \oint_{S^1_\infty} \, 
(d z \p_z - d \bar z \p_{\bar z}) \log \det (H_0 H_0^\dagger),
\eeq
where $S^1_{\infty}$ denotes the clock-wise circle 
at infinity $|z| \rightarrow \infty$.
This equation shows that the order of the polynomial $\det H_0(z)$ 
corresponds to the vorticity. 

As an example, let us consider 
the case of single vortex configurations. 
The simplest single vortex solution can be obtained 
by embedding the single Abelian vortex solution in 
the upper-left corner of the $N$-by-$N$ matrices $H$ and $W_{\bar z}$
\beq
H ~=~ \ba{cc} H^{\rm ANO} & 0 \\ 0 & v \, \mathbf 1_{N-1} \ea, \hs{10}
W_{\bar z} ~=~ \ba{cc} W_{\bar z}^{\rm ANO} & 0 \\ 0 & \mathbf 0_{N-1} \ea. 
\eeq
The other single vortex solutions can be obtained by 
acting on the embedded solution with the $SU(N)_{\rm C+F}$ symmetry
\beq
H \rightarrow U^\dagger H U, \hs{10} W_{\bar z} \rightarrow U^\dagger W_{\bar z} U, \hs{10} U \in SU(N)_{\rm C + F}.
\eeq
Since the vortex embedded into the upper-left corner breaks 
$SU(N)_{\rm C+F} \rightarrow SU(N-1) \times U(1)$, 
the internal vortex moduli space is the complex projective space
\beq
\C P^{N-1} \cong \frac{SU(N)}{SU(N-1) \times U(1)}.
\eeq 
By modding out the unbroken $SU(N-1) \times U(1)$, 
we can fix the unitary matrix $U$ as
\beq
U &=& \ba{cc} 1 & - \vec b^\dagger \\ 0 & \mathbf 1_{N-1} \ea 
\ba{cc} X^{\frac{1}{2}} & 0 \\ 0 & Y^{-\frac{1}{2}} \ea 
\ba{cc} 1 & 0 \\ \vec b & \mathbf 1_{N-1} \ea,
\eeq
where the parameters $\vec b = (b^1, \cdots, b^{N-1})$ 
can be interpreted as the inhomogeneous coordinates of $\C P^{N-1}$ and 
$X$ and $Y$ are given by\footnote{The square root of the matrix $Y$ is given by $Y^{\pm \frac{1}{2}} = (1-P) + (1+|\vec b|^2)^{\pm \frac{1}{2}} P$ with $P = |\vec b|^{-2} \vec b \otimes \vec b^\dagger$.}
\beq
X \equiv 1 + \vec b^\dagger \cdot \vec b, \hs{10} 
Y \equiv \mathbf 1_{N-1} + \vec b \otimes \vec b^\dagger.
\eeq
The physical meaning of the orientational moduli can be seen 
from the magnetic flux
\beq
F_{z \bar z} ~=~ F_{z \bar z}^{\rm ANO} ~\times~ \frac{1}{1+|\vec b|^2} \ba{cc} \ \, 1 & - \vec b^\dagger \\ -\vec b & \ \vec b \otimes \vec b^\dagger \ea.
\eeq
Thus the orientational moduli $\vec b$ determines 
the $U(1)$ generator in which the vortex magnetic flux is embedded.

The moduli matrix $H_0$ and 
the solution of the master equation $\Omega$ 
corresponding to the single vortex configurations are given by
\beq
H_0(z) ~=~ 
V \ba{cc} z - Z & 0 \\ 0 & \mathbf 1_{N-1} \ea U, \hs{10}
\Omega ~=~ V \ba{cc} e^{\psi} & 0 \\ 0 & \mathbf 1_{N-1} \ea V^\dagger,
\label{eq:single_sol}
\eeq
where $\psi$ is a real profile function satisfying
\beq
\frac{4}{g^2 v^2} \p_{\bar z} \p_z \psi = 1 - |z-Z|^2 e^{-\psi},
\eeq
with the boundary condition $\psi \rightarrow \log |z-Z|^2$. 
The complex parameter $Z$ in the moduli matrix $H_0$ 
can be interpreted as the position moduli of the vortex. 
Although the holomorphic matrix $V(z)$ can be 
an arbitrary element of $GL(N,\C)$, 
it will be convenient to use the matrix of the form
\beq
V(z) &=& 
\ba{cc} X^{-\frac{1}{2}} & 0 \\ 0 & Y^{\frac{1}{2}} \ea
\ba{cc} 1 & (z-Z) \vec b^\dagger \\ 0 & \mathbf 1_{N-1} \ea.\label{eq:SUtrf-V}
\eeq
This $V$-transformation is chosen so that the moduli matrix 
$H_0(z)$ takes the form
\beq
H_0(z) ~=~ \ba{cc} z - Z & 0 \\ \vec b & \mathbf 1_{N-1} \ea.
\eeq
The important point is that the $V$-transformation is completely fixed 
so that the moduli matrix is holomorphic not only in $z$ 
but also in the moduli parameters $Z$ and $\vec b$. 
In general, there exists such a fixed form of the moduli matrix 
in each coordinate patch of the moduli space. 
The $V$-transformation between a pair of the fixed moduli matrices 
induces a coordinate transformation from one patch to another. 
For instance, in the case of $N=2$, there are two fixed forms 
\beq
H_0 = \ba{cc} z - Z & 0 \\ b & 1 \ea, \hs{10} H_0' = \ba{cc} 1 & b' \\ 0 & z - Z \ea. 
\eeq
These two matrices are related by 
\begin{eqnarray}
H_0'= V H_0 , \hs{10} V = \ba{cc} 0 & b'\\ -b & z-Z \ea, \hs{5} 
b'=\frac{1}{b}. 
\end{eqnarray}
Thus, the induced coordinate transformation is 
the standard transition map between the 
inhomogeneous coordinates of $\C P^1$. 
As this example shows, the coordinate fransformations are holomorphic, 
so that the moduli space is a complex manifold. 

For a general winding number $k$, 
we can fix generic moduli matrices to the following form 
\cite{Eto:2005yh}
\beq
H_0(z) = 
{\arraycolsep 3mm
{\renewcommand{\arraystretch}{1.2}
\ba{cc} (z-Z_1) (z-Z_2) \cdots (z-Z_k) & 0 \\ \vec b_1 \, e_1(z) + \vec b_2 \, e_2(z) \cdots + \vec b_k \, e_k(z) & \mathbf 1_{N-1} \ea}}, \hs{7}
e_I(z) = \prod_{J \not = I} \frac{z-Z_J}{Z_I-Z_J}.\label{eq:genericH0}
\eeq
The parameters $Z_I$ and $\vec b_I~(I=1,\cdots,k)$ 
are position and orientational moduli of $I$-th vortex 
which cover a local coordinate patch 
of the moduli space of the vortices $\mathcal M_k$. 
As in the case of the single vortex, 
the moduli matrix $H_0(z)$ is holomorphic 
with respect to the moduli parameters. 
This fact will be important 
when we derive general formulas for the second and 
fourth order effective Lagrangian in section \ref{sec:deriv}.
%\ko
%{There we will see that 
%the holomorphic $V$-transformation causing transition between
%coordinate patches keeps those formulas invariant whereas generic
%$V$-transformation like  Eq.\eqref{eq:SUtrf-V} does not.}
%%%%%%%%%%%%%%%%%%%%%%%%%%%%%%%%%%%%%%%%%%%%%%%%%%%%%%%%%%%%%%%%%%%%%%%%%%%
%\subsection{The second order effective action}
%The leading order corrections
%have been well studied for various solitons 
%\cite{Manton:1981mp,Atiyah:1985dv,Atiyah:1988jp,Gibbons:1995yw,Lee:1996kz,
%Taubes:1979tm,Samols:1991ne,Chen:2004xu,Manton:2002wb,Krusch:2009tn}
%and it gives a natural metric on the moduli space manifold.
%The soliton dynamics of the leading order is then identified with the geodesic %motion of the moduli space. 
%In what follows we will study higher derivative corrections 
%to the leading order effective Lagrangian by following 
%the prescriptions explained in the previous section.

%%%%%%%%%%%%%%%%%%%%%%%%%%%%%%%%%%%%%%%%%%%%%%%%%%%%%%%%%%%%%%%%%%%%%%%%%%%
\section{Preliminary: a particle in $\mathbb{R}^n$}\label{sec:prelim}

Before studying higher derivative corrections to 
the vortex effective action in field theories, 
we first explain our basic strategy in a simple system of 
a particle in $\mathbb{R}^n$ with the following Lagrangian
\beq
L ~=~ \frac{m}{2} \dot{\bf x} \cdot \dot{\bf x} - V({\bf x}), \hs{10} 
{\bf x}(t) = (x^1,x^2,\cdots,x^n).
\label{eq:lag_particle}
\eeq
In a minimum energy configuration, the particle stays 
at the bottom of the potential $V$ where the gradient of $V$ vanishes
\beq
{\rm grad} \, V = 0. 
\eeq
Now let us assume that the potential $V$ has flat directions 
which are parameterized by $\phi^i$. 
Then we can define ``the moduli space'' of the minimum energy configurations by
\beq
{\cal M} = \left\{ \ {\bf x}^{(0)}(\phi^i) \in \mathbb{R}^n \ |\ {\rm grad} \, V = 0\ \right\},
\eeq
and $\phi^i$ can be interpreted as ``the moduli parameters''. 
Since one can shift the particle to any points 
on the bottom of the potential without loss of energy, 
we can assume that the particle slowly moves 
along the moduli space $\mathcal M$ 
for sufficiently small excitation energy.
This motion of the particle can be represented by 
the moduli parameters $\phi^i(t)$ which weakly depend on the time $t$, 
that is 
\beq
{\bf x}(t) = {\bf x}^{(0)}(\phi^i(t)).
\label{eq:slow_x}
\eeq
Then, the low energy effective Lagrangian of the particle can be obtained 
by substituting Eq.\,\eqref{eq:slow_x} 
into the original Lagrangian Eq.\,\eqref{eq:lag_particle}
\beq
L_{\rm eff} ~=~ L_{\rm eff}^{(0)} + L_{\rm eff}^{(2)} ~= - V({\bf x}^{(0)}) + \frac{m}{2} \dot{\bf x}^{(0)} \cdot \dot{\bf x}^{(0)},
\label{eq:2nd}
\eeq
where $L_{\rm eff}^{(0)} = - V({\bf x}^{(0)})$ is 
the constant value of the potential at the bottom. 
The second order effective Lagrangian 
$L_{\rm eff}^{(2)}$ can be rewritten by using 
the metric $g_{ij}$ on the moduli space $\mathcal M$ as
\beq
L_{\rm eff}^{(2)} ~=~ \frac{m}{2} g_{ij} \dot\phi^i\dot\phi^j.
\eeq
The metric is given by the inner products of 
the basis $\{ \boldsymbol \Phi_i \}$ of ``the zero modes"
\beq
g_{ij} \equiv \boldsymbol \Phi_i \cdot \boldsymbol \Phi_j, \hs{10}
\boldsymbol \Phi_i ~\equiv~ \frac{\p}{\p \phi^i} {\bf x}^{(0)}. 
\label{eq:gij}
\eeq
The effective equations of motion for the moduli parameters 
take the form of the geodesic equation
\beq
\ddot \phi^i + \Gamma^i_{jk} \dot\phi^j \dot\phi^k ~=~ 0,
\label{eq:second_eom}
\eeq
where the connection is given by
\beq
\Gamma^i_{jk} ~=~ \frac{1}{2} g^{il} \left(g_{lj,k}+g_{lk,j}-g_{jk,l}\right) ~=~g^{il} \, \boldsymbol \Phi_l \cdot \frac{\p}{\p \phi^j} \boldsymbol \Phi_k.
\label{eq:connection}
\eeq
This approximation is valid if the velocity of the particle is 
sufficiently small so that the time derivative is much less 
than the typical mass scale of the massive modes determined 
from the Hessian matrix $\mathbf H$ of the potential $V$ defined by 
\beq
\big[ \mathbf H ({\bf x}^{(0)}) \big]_{ab} = \frac{\p^2}{\p x^a \p x^b} V \bigg|_{{\bf x}={\bf x}^{(0)}}.
\eeq

Next, let us consider higher derivative corrections 
to the effective Lagrangian 
by taking the massive modes into account. 
To this end, we first introduce a correction to \eqref{eq:slow_x}
by adding small fluctuations to 
``the slowly moving background" ${\bf x}^{(0)}(\phi^i(t))$ as
\beq
{\bf x}(t) ~=~ {\bf x}^{(0)}(\phi^i(t)) + \delta {\bf x}(t).
\eeq
Since the motion of the particle along the flat direction is 
already represented by the moduli parameters $\phi^i(t)$, 
we impose the following condition 
to avoid the double-counting of the degrees of freedom of the zero modes:
\beq
\delta {\bf x} \cdot \boldsymbol \Phi_i = 0.
\label{eq:orthogonal}
\eeq
This means that the fluctuations $\delta {\bf x}$ 
is orthogonal to the tangent space of ${\cal M}$, 
namely {\it the fluctuation $\delta {\bf x}$ contains only massive modes}. 
Then, the original Lagrangian Eq.\,\eqref{eq:lag_particle} can be rewritten as
\beq
L &=& L_{\rm eff}^{(0)} + L_{\rm eff}^{(2)} + \delta L + \lambda^i \left( \delta {\bf x} \cdot \boldsymbol \Phi_i \right) , \phantom{\bigg[}
\label{eq:L_higher}\\
\delta L &=& 
m \, \dot{\bf x}^{(0)} \cdot \delta \dot{\bf x} 
+ \frac{m}{2} \delta \dot{\bf x} \cdot \delta \dot{\bf x} 
- \frac{1}{2} \delta {\bf x} \, \mathbf H \, \delta {\bf x} + \cdots, \phantom{\bigg[}
\label{eq:l4}
\eeq
where we have introduced the Lagrange multipliers $\lambda^i$ 
to impose the constraint Eq.\,\eqref{eq:orthogonal}. 
Note that there is no linear term in the Taylor expansion of the potential 
since the background satisfies ${\rm grad} \, V( {\bf x}^{(0)} ) = 0$. 
Now let us consider the expansion of $\delta {\bf x}$ with respect to the time derivative $\p_t$ 
\beq
\delta {\bf x} = {\bf x}^{(2)} + {\bf x}^{(4)} + \cdots, \hs{10}
{\bf x}^{(n)} \sim \mathcal O(\p_t^n). 
\eeq
Note that $\delta {\bf x}$ can have only terms 
with even numbers of the time derivatives 
due to the reflection symmetry $t \rightarrow - t$. 
There is no zeroth order term  
since the fluctuation $\delta {\bf x}$ vanishes for the static configurations.
Correspondingly, the Lagrangian $\delta L$ can also be expanded as
\beq
\delta L = L^{(4)} + L^{(6)} + \cdots .
\eeq
The fourth order Lagrangian contains the background ${\bf x}^{(0)}$ 
and the second order fluctuation ${\bf x}^{(2)}$
\beq
L^{(4)} = m \, \dot{\bf x}^{(0)} \cdot \dot{\bf x}^{(2)}
- \frac{1}{2} {\bf x}^{(2)} \, \mathbf H \, {\bf x}^{(2)}.
\label{eq:l4-2}
\eeq
Note that $\mathbf x^{(4)}$ does not contribute to the fourth order Lagrangian
since ${\rm grad} \, V( {\bf x}^{(0)} ) = 0$.
The massive modes ${\bf x}^{(2)}$ can be eliminated 
by solving their equation of motion
\beq
\mathbf H \, {\bf x}^{(2)} + m \, \ddot{\bf x}^{(0)} = \lambda^i \boldsymbol \Phi_i.
\label{eq:massive_eq}
\eeq
To determine the Lagrange multiplier $\lambda^i$, 
let us take the inner products of the both hand sides and
the zero modes $\boldsymbol \Phi_j$
\beq
m \, \boldsymbol \Phi_j \cdot \ddot{\bf x}^{(0)} &=& g_{ij} \lambda^i,
\eeq
where we have used the fact that the zero modes 
$\left\{ \boldsymbol \Phi_i \right\}$ satisfy
\beq
\boldsymbol \Phi_i \, \mathbf H ~=~ \frac{\p}{\p \phi^i} \big[ \, {\rm grad} \, V({\bf x}^{(0)}) \, \big] ~=~ 0.
\label{eq:zero_mode}
\eeq
We can show that the Lagrange multiplier is proportional to 
the second order equation of motion Eq.\,\eqref{eq:second_eom}
\beq
\lambda^i ~=~ m \, g^{ij} \, \boldsymbol \Phi_j \cdot \ddot{\bf x}^{(0)} ~=~ m \left( \ddot \phi^i + \Gamma^i_{kl} \dot \phi^k \dot \phi^l \right).
\eeq
Substituting back $\lambda^i$ into Eq.\,\eqref{eq:massive_eq}, 
we obtain the following equations of motion for the massive modes
\beq
\mathbf H \, {\bf x}^{(2)} = - m \, \mathbf P \ddot{\bf x}^{(0)},
\label{eq:massve_eom}
\eeq
where $\mathbf P$ is the projection matrix 
which project out the zero modes
\beq
\mathbf P \ddot{\bf x}^{(0)} ~\equiv~ \ddot{\bf x}^{(0)} - g^{ij} (\ddot{\bf x}^{(0)} \cdot \boldsymbol \Phi_j ) \boldsymbol \Phi_i, 
%\delta^{ab} - g^{ij} \frac{\p x^a}{\p \phi^i} \frac{\p x^b}{\p \phi^j}.
\eeq
Since the right hand side of Eq.\,\eqref{eq:massve_eom} does not 
contain the zero mode directions, 
the matrix $\mathbf H$ can be ``inverted" as
\beq
{\bf x}^{(2)} = - m \, \mathbf G \ddot{\bf x}^{(0)}, 
\label{eq:massive_sol}
\eeq
where $\mathbf G$ is the matrix satisfying
\beq
\mathbf H \mathbf G = \mathbf P, \hs{10} 
\mathbf G \mathbf P = \mathbf P \mathbf G = \mathbf G, \hs{10}
\mathbf G \boldsymbol \Phi_i = 0. 
\eeq
Substituting the solution for the massive modes Eq.\,\eqref{eq:massive_sol} 
into Eq.\,\eqref{eq:l4-2}, 
we obtain the following fourth order effective Lagrangian
\beq
L_{\rm eff}^{(4)} ~=~ \frac{1}{2} {\bf x}^{(2)} \mathbf H \, {\bf x}^{(2)} 
~=~ \frac{m^2}{2} \dot \phi^i \dot \phi^j \dot \phi^k \dot \phi^l \left( \frac{\p \boldsymbol \Phi_j}{\p \phi^i} \mathbf G \frac{\p \boldsymbol \Phi_l}{\p \phi^k} \right).
\label{eq:4th-L}
\eeq

\begin{figure}[h]
\begin{center}
\includegraphics[width=60mm]{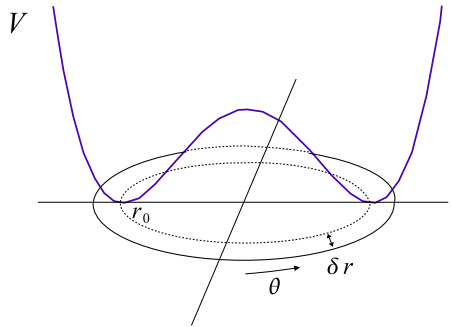}
\caption{{\footnotesize The shift of the rotation radius.}}
\label{fig:shift}
\end{center}
\end{figure}

Let us see a simple example of a particle in $\mathbb{R}^2$.
We assume that the potential is rotationally symmetric 
$V = V(|{\bf x}|)$ and has a minimum at $|{\bf x}| = r_0$, 
namely $V'(r_0)=0$ and $V''(r_0) > 0$. 
The static configurations are parameterized by the moduli 
parameter $\theta$ as 
\beq
{\bf x}^{(0)} ~= \ba{cc} r_0 \cos \theta \\ r_0 \sin \theta \ea .
\eeq
The second order effective Lagrangian is given by
\beq
L_{\rm eff}^{(2)} ~=~ \frac{m}{2} \dot{\bf x}^{(0)} \cdot \dot{\bf x}^{(0)} ~=~ \frac{m}{2} r_0^2 \, \dot \theta^2.\label{eq:2ndEL}
\eeq
The corresponding equation of motion 
for the moduli parameter is $\ddot \theta = 0$ and 
describes the particle rotating around the circle at $|{\bf x}| = r_0$.
The matrices $\mathbf H$ and $\mathbf G$ are respectively given by
\beq
\mathbf H ~=~ V''(r_0) \ \mathbf P, \hs{10} 
\mathbf G ~=~ V''(r_0)^{-1} \mathbf P,
\eeq
where $\mathbf P$ is the projection operator
\beq
\mathbf P ~= \ba{cc} \cos^2 \theta & \sin \theta \cos \theta \\ \sin \theta \cos \theta & \sin^2 \theta \ea.
\eeq
From Eq.\,\eqref{eq:4th-L}, 
we obtain the following fourth order Lagrangian
\beq
L_{\rm eff}^{(4)} ~=~ \frac{(m r_0)^2}{2} V''(r_0)^{-1} \, \dot \theta^4. 
\eeq
Even if we take into account this fourth order Lagrangian, 
the equation of motion is not modified $\ddot \theta=0$
and solved by $\theta = \omega t + \theta_0$. 
On the other hand, 
the relation between the angular velocity $\omega$ 
and the angular momentum $l$ is modified as
\beq
l ~=~ m r_0^2 \left[ 1 + 2 m \omega^2 V''(r_0)^{-1} \right] \omega.
\label{eq:angular}
\eeq
This is just because the rotation radius 
is increased by the centrifugal force as 
\beq
r_0 ~~\longrightarrow~~ r_0 + m r_0 \, \dot \theta^2 \, V''(r_0)^{-1}.
\eeq
The shift of the rotation radius can be 
also seen in the solution for the massive mode 
\beq
{\bf x}^{(0)} + \delta {\bf x} ~\approx~ {\bf x}^{(0)} - m \, \mathbf G \ddot{\bf x}^{(0)} ~=~ \left[ r_0 + m r_0 \dot \theta^2 \, V''(r_0)^{-1} \right] \ba{cc} \cos \theta \\ \sin \theta \ea.
\eeq
Therefore, the higher derivative term 
gives the correction from the massive mode 
which is slightly shifted by the motion of the zero mode 
(see Fig.\,\ref{fig:shift}). 

%%%%%%%%%%%%%%%%%%%%%%%%%%%%%%%%%%%%%%%%%%%%%%%%%%%%%%%%%%%%%%%%
\section{The effective theory \label{sec:deriv}}
\subsection{Derivative expansion}
In this section, we discuss the effective Lagrangian for non-Abelian vortices 
by generalizing the method of the derivative expansion 
discussed in the previous section. 
To deal with the vortex positions and orientations 
$Z_I,\,\vec b_I~(I=1,\cdots,k)$ on an equal footing, 
we combine them into a set of complex moduli parameters 
$\phi^i~(i=1,\cdots,k N = {\rm dim}_\C \, \mathcal M_k)$, 
and assume that the moduli matrix $H_0(z)$ 
is always holomorphic in $\phi^i$ like Eq.\,\eqref{eq:genericH0}.

The zero mode fluctuations along the vortex world-volume are 
described by the vortex moduli $\phi^i$ 
promoted to fields which weakly depend on the world-volume coordinates
\beq
\phi^i \rightarrow \phi^i(x^\alpha), \hs{10} (\alpha = 0,1,\cdots,d-2).
\label{eq:promotion}
\eeq
These moduli fields induce fluctuations of massive modes
around the weakly fluctuating vortex background 
\beq
H (x^\mu)~ &=& H^{(0)} \, (z,\bar z,\phi(x^\alpha)) ~+~ \delta H(x^\mu), \\
W_{\bar z} (x^\mu) &=& W_{\bar z}^{(0)}(z,\bar z,\phi(x^\alpha)) ~+~ \delta W_{\bar z}(x^\mu), \\
W_\alpha (x^\mu) &=& \hs{12} 0 \hs{16} ~+~ \delta W_\alpha (x^\mu), 
\label{eq:fluctuations}
\eeq
where $H^{(0)}$ and $W_{\bar z}^{(0)}$ are the BPS vortex background 
Eq.\,\eqref{eq:BPSsol} depending on the world-volume coordinates $x^\alpha$ 
through the moduli fields $\phi(x^\alpha)$. 
We assume that the excitation energy of the fluctuations are much less than 
the typical mass scale of the massive modes. 
Then we can expand the induced fields  
with respect to the derivative by assuming that
\beq
\p_\alpha ~\ll~ g v.
\eeq
Note that $g v$ is the unique mass scale 
controlling the mass of the bulk fields, inverse width of the vortex 
and the mass scale of the massive modes localized on the vortex world-volume. 
The induced fluctuations are expanded 
with respect to the derivative $\partial_\alpha$ as 
\beq
\delta H~ &=& H^{(2)} + H^{(4)} + \cdots, \\
\delta W_{\bar z} &=& W_{\bar z}^{(2)} + W_{\bar z}^{(4)} + \cdots, \\
\delta W_\alpha &=& W_\alpha^{(1)} + W_\alpha^{(3)} + \cdots.
\eeq
These fluctuations can be determined by solving the equations of motion 
(\ref{eq:eom_full1}) and (\ref{eq:eom_full2}) order-by-order. 
Note that odd (even) order equations of motion 
for $\delta H$ and $\delta W_{\bar z}$ ($\delta W_\alpha$) are trivial
due to the reflection symmetry $x^\alpha \rightarrow - x^\alpha$. 
Then, the effective Lagrangian can be obtained 
by eliminating the induced fluctuations from original Lagrangian 
\eqref{eq:action} expanded with respect to the derivative $\partial_\alpha$
\beq
\mathcal L ~=~ \mathcal L^{(0)} + \mathcal L^{(2)} + \mathcal L^{(4)} + \cdots. 
\eeq
As we will see, the Lagrangian is quadratic in
the induced fluctuations up to the fourth order in the derivative $\p_\alpha$, 
so that we can use the linearized equations of motion 
to eliminate the induced fluctuations.
In order to determine the fluctuations, 
we have to specify the boundary conditions for them. 
In the singular gauge, the vortex background takes the following form 
at the infinity $|z| \rightarrow \infty$
\beq
H^{(0)} ~=~ v \mathbf 1_N + \mathcal O(e^{-g v |z|}), 
\hs{10}
W_{\bar z}^{(0)} ~=~ \mathcal O ( \bar z^{-1} ).
\eeq
Since this vortex background is independent of 
the zero mode fluctuations $\phi(x^\alpha)$ at infinity, 
the minimal excitations induced 
by the zero modes should vanish at infinity.
Therefore, we impose the following boundary conditions 
for the induced fluctuations
\beq
\delta H \underset{|z| \rightarrow \infty}{\longrightarrow} 0 , \hs{10} 
\delta W_\mu \underset{|z| \rightarrow \infty}{\longrightarrow} 0, \hs{10} (\mbox{in the singular gauge}).
\label{eq:boundary}
\eeq

%%%%%%%%%%%%%%%%%%%%%%%%%%%%%%%%%%%%%%%%%%%%%%%%%%%%%%%%%%%%%%%%%%%%%%%%%%%
\subsection{General formulas for the effective Lagrangian}

%%%%%%%%%%%%%%%%%%%%%%%%%%%%%%%%%%%%%%%%%%%%%%%%%%%%%%%%%%%%%%%%%%%%%%%
\subsubsection{The zeroth order effective Lagrangian}
The zeroth order Lagrangian contains only the background fields
\beq
\mathcal L^{(0)} = \tr \left[ - \frac{4}{g^2} |F_{z \bar z}^{(0)}|^2
- 2 \D_z H^{(0)} (\D_z H^{(0)})^\dagger - \frac{g^2}{4} (H^{(0)}H^{(0)\dagger} - v^2 \mathbf 1_N)^2 \right].
\eeq
Substituting the solution \eqref{eq:BPSsol} and 
using the boundary condition Eq.\,\eqref{eq:boundary_Omega}, 
we find that the zeroth order term 
%\meto
{$\mathcal L_{\rm eff}^{(0)}$ of the low-energy effective Lagrangian} 
is given 
by the sum of the tension of the individual vortices
\beq
\mathcal L_{\rm eff}^{(0)} 
~\equiv~ \int d^2 x \, \mathcal L^{(0)} 
= - k T, \hs{10} T ~\equiv~ 2 \pi v^2.
\eeq

%%%%%%%%%%%%%%%%%%%%%%%%%%%%%%%%%%%%%%%%%%%%%%%%%%%%%%%%%%%%%%%%%%%%%%%
\subsubsection{The second order effective Lagrangian}

The second order Lagrangian takes the form \cite{Hanany:2003hp,Auzzi:2003fs} 
(see also \cite{Shifman:2004dr,Isozumi:2004vg,Eto:2004rz,Eto:2005yh}).
\beq
\mathcal L^{(2)} &=& \tr \left[ 
\frac{4}{g^2}
F_{\alpha \bar z}^{(1)} {F^\alpha}_{z}^{(1)}
+ \D_\alpha H^{(0)} (\D^\alpha H^{(0)})^\dagger \right].
\label{eq:secondL}
\eeq
where $F_{\alpha \bar z}^{(1)}$ and $\D_\alpha H^{(0)}$ are given by
\beq
F_{\alpha \bar z}^{(1)} ~=~ \p_\alpha W_{\bar z}^{(0)} - \p_{\bar z} W_\alpha^{(1)} + i [W_\alpha^{(1)}, W_{\bar z}^{(0)}], \hs{5}
\D_\alpha H^{(0)} ~=~ \p_\alpha H^{(0)} + i W_\alpha^{(1)} H^{(0)}.
\eeq
Note that the terms containing 
the second order fields $H^{(2)}$ and $W_{\bar z}^{(2)}$ 
do not contribute to the second order Lagrangian 
since these terms are proportional to the background equations of motion
\beq
0 ~=~ \tr \left[ H^{(2)} \left( \frac{\delta}{\delta H} \int d^{d+1} x \, \mathcal L^{(0)} \right)_{\rm BPS \ background} \right], ~~\cdots. 
\label{eq:vanishingH2term}
\eeq
The dynamical degrees of freedom in the second order Lagrangian are
not only the zero modes $\phi^i(x^\alpha)$ contained in the
background fields ($H^{(0)}$, $W_{\bar z}^{(0)}$) 
but also the first order fluctuations $W_\alpha^{(1)}$.
We can eliminate $W_\alpha^{(1)}$ by using the equations of motion\footnote{
Note that the orthogonality condition between zero modes and fluctuations is trivial for $W_\alpha^{(1)}$ since $W_\alpha = 0$ in the vortex background.}
\beq
\frac{4}{g^2} \left( \D_z F_{\bar z \alpha}^{(1)} + \D_{\bar z} F_{z \alpha}^{(1)} \right) ~=~ i \Big[ H^{(0)} (\D_\alpha H^{(0)})^\dagger - \D_\alpha H^{(0)} H^{(0)\dagger} \Big].\label{eq:W1}
\eeq
By using the fact that 
the moduli matrix $H_0(z)$ is holomorphic in the moduli parameters $\phi^i$ 
and the matrix $\Omega$ is the solution of 
the master equation Eq.\,\eqref{eq:master}, 
we can check that the solution is given by
\beq
W_\alpha^{(1)}
~=~ i (\delta_\alpha S^\dagger S^{\dagger-1} - S^{-1} \delta_\alpha^\dagger S),
\label{eq:Firstsol}
\eeq
where we have defined the differential operators 
$\delta_\alpha$ and $\delta_\alpha^\dagger$ by
\beq
\delta_\alpha \equiv \p_\alpha \phi^i \frac{\p}{\p \phi^i}, \hs{10}
\delta_\alpha^\dagger \equiv \p_\alpha \bar \phi^i \frac{\p}{\p \bar \phi^i}.
\eeq
The solution Eq.\,\eqref{eq:Firstsol} satisfies 
the boundary condition Eq.\,\eqref{eq:boundary} 
since the asymptotic form of the matrix $S$ in the singular gauge
is $S \rightarrow H_0(z,\phi^i)$ and 
the moduli matrix $H_0(z,\phi^i)$ is holomorphic with respect to
$\phi^i$.
Substituting the solution \eqref{eq:Firstsol} 
into the second order Lagrangian \eqref{eq:secondL}, 
we obtain the following formal expression of 
the second order effective Lagrangian
\beq
\mathcal L_{\rm eff}^{(2)} ~=~ v^2
\int d^2 x \ \delta_\alpha^\dagger \tr \left[ \delta_\alpha H_0
H_0^\dagger \Omega^{-1} \right] ~=~ g_{i \bar j} \p_\alpha \phi^i
\p^\alpha \bar \phi^j. \label{eq:2nd}
\eeq
This effective Lagrangian gives a natural K\"ahler metric on
the moduli space of vortices
\beq
g_{i \bar j} ~\equiv~ v^2 \int d^2 x \, \frac{\p}{\p \bar \phi^j} \tr \left[ \frac{\p H_0}{\p \phi^i}
 H_0^\dagger \Omega^{-1} \right].
\label{eq:metric}
\eeq
Note that the above formulas for 
the first order solution \eqref{eq:Firstsol} 
and the moduli space metric 
are not invariant under the generic $V$-transformation Eq.\eqref{eq:V-trf}. 
However they are invariant under the $V$-transformation 
respecting the holomorphy of $H_0(z)$ 
as we pointed out at the end of Sec.\,\ref{sec:NAvor}. 

In order to obtain the explicit form of the moduli space metric, 
one needs to know the solution of the master equation $\Omega$.
However, no analytic solution of 
$\Omega$ has been known even for the minimal winding vortex, 
and hence it is quite difficult to obtain 
the explicit form of the effective Lagrangian in general. 
Nevertheless, we can obtain 
an exact form of the second order effective Lagrangian 
for the single vortex since the degrees of freedom $Z,\,\vec b$ in this case 
are nothing but Nambu-Goldstone zero modes \cite{Eto:2004rz}. 
By substituting the solution Eq.\,\eqref{eq:single_sol} 
and using the boundary condition for the profile function 
$\psi \rightarrow \log |z-Z|^2$, we can show that
\beq
\mathcal L^{(2)}_{\rm eff} ~=~ \frac{T}{2} \p_\alpha Z \p^\alpha \bar Z + \frac{4\pi}{g^2} g_{i \bar j}^{\rm FS} \p^\alpha b^i \p_\alpha \bar b^j , 
\eeq
where $g_{i \bar j}^{\rm FS}$ is the Fubini-Study metric on $\C P^{N-1}$
\beq
g_{i \bar j}^{\rm FS} ~\equiv~ \frac{\p}{\p b^i} \frac{\p}{\p \bar b^j} \log (1 + |\vec b|^2).
\eeq
Another example of analytic moduli space metric for higher 
winding vortices has been obtained for well-separated 
vortices \cite{Fujimori:2010fk}.

Note the moduli space metric can be 
rewritten as inner products of zero modes 
as in the case of the example discussed 
in the previous section (see Eq.\,\eqref{eq:gij}). 
Before closing this subsection, 
let us review the zero modes 
and the derivative operator defining them \cite{Hanany:2003hp}.
The physical zero modes in a BPS background 
are defined as the solutions of the linearized BPS equations 
for the fluctuations around the background. 
With a suitable gauge fixing condition, 
the linearized BPS equations can be written as
\beq
\Delta \Phi ~\equiv~
{\renewcommand{\arraystretch}{1.2}
\ba{cc} 
i \D_{\bar z}^f & - \frac{g}{2} H_r^{(0)} \\ 
\frac{g}{2} H_r^{(0)\dagger} & i \D_z^a \ea 
\ba{cc} \delta H \\ \frac{2}{g} \delta W_{\bar z} \ea} ~=~ 0,
\eeq
where the subscript $r$ denotes the fact that $H^{(0)}$ 
acts as right multiplication and 
$\D_{\bar z}^f$ and $\D_z^a$ are covariant derivatives 
with the zeroth order gauge field 
which act on the fundamental and adjoint fields, respectively.
There exists one zero mode $\Phi_i$ for each moduli parameter $\phi^i$ 
and the set $\{\Phi_i\}~(i=1,\cdots,{\rm dim}_\C \, \mathcal M)$ 
forms a basis of the zero modes\footnote{
The linearized equations of motion in a static BPS background 
can be summarized into the form of $\Delta^\dagger \Delta \Phi = 0$. 
Here, the Hermitian conjugate of the operator $\Delta$ 
\begin{eqnarray}
\Delta^\dagger~=~ 
\ba{cc} 
i \D_{ z}^f & \frac{g}{2} H_r^{(0)} \\ 
-\frac{g}{2} H_r^{(0)\dagger} & i \D_{\bar z}^a 
\ea, \notag
\end{eqnarray} 
turns out to have no zero mode \cite{Hanany:2003hp}; 
$\Delta^\dagger \Phi = 0 \, \Rightarrow \, \Phi=0$,
so that $\{\Phi_i\}$ gives a full set of the zero modes. 
}. 
In terms of $\Omega, S$ and $H_0$, 
the basis $\Phi_i$ can be written as
\begin{eqnarray}
\Phi_i~=~{\renewcommand{\arraystretch}{1.5} \ba{c} 
vS^{-1} \Omega \, \displaystyle \frac{\p}{\p \phi^i} 
 \left[ \Omega^{-1} H_0 \right] \\ 
\displaystyle \frac{2i}{g}S^{-1} \bar \p \left[ \Omega \displaystyle 
\frac{\p}{\p \phi^i} \Omega^{-1} \right] S \ea}.
\end{eqnarray}
By using the solution for $W_\alpha^{(1)}$ given in Eq.\,\eqref{eq:Firstsol}, 
these zero modes can be summarized as
\beq
\p_\alpha \phi^i \, \Phi_i ~=~ \ba{c} \D_\alpha H^{(0)} \\ 
\frac{2}{g} F_{\alpha \bar z}^{(1)} \ea.
\eeq
Then, the moduli space metric Eq.\,\eqref{eq:metric} 
can be rewritten as inner products of the physical zero modes $\Phi_i$ 
\beq
g_{i \bar j} \equiv \big< \Phi_j \,,\, \Phi_i \big>.
\eeq
where we have defined the hermitian inner product for pairs of fields 
in the fundamental and adjoint representations 
\beq
\big< \Phi , \Phi' \big> ~\equiv~ \int d^2 x \, \tr\left[ f' f^\dagger +  a^\dagger a' \right], \hs{10} \Phi = \ba{c} f \\ a \ea, ~~ \Phi' = \ba{c} f' \\ a' \ea.
\eeq
%%%%%%%%%%%%%%%%%%%%%%%%%%%%%%%%%%%%%%%%%%%%%%%%%%%%%
\subsubsection{The fourth order effective Lagrangian}
Let us now calculate the fourth order effective Lagrangian 
by solving the linearized equations of motion 
and eliminating the massive modes. 
The fourth order Lagrangian takes the form
\beq
\mathcal L^{(4)} \hs{-2} &=& \hs{-2} \tr \Bigg[ - \frac{1}{2g^2} (F_{\alpha \beta}^{(2)})^2 + \left( \frac{4}{g^2} \D^\alpha W_{\bar z}^{(2)} F_{\alpha z}^{(1)} + \D_\alpha H^{(2)} ( \D^\alpha H^{(0)} )^\dagger + (h.c.) \right) \notag \\
&{}& ~~~ \, - \frac{4}{g^2} \left| i \D_z W_{\bar z}^{(2)} + \frac{g^2}{4} H^{(2)} H^{(0)\dagger} + (h.c.) \right|^2 - 4 \left| \D_{\bar z} H^{(2)} + i W_{\bar z}^{(2)} H^{(0)} \right|^2 \Bigg].
\eeq
Here, terms proportional to 
$W_\alpha^{(3)},H^{(4)}$ and $,W_z^{(4)}$ automatically vanish 
due to the same mechanism as Eq.\eqref{eq:vanishingH2term}. 
It will be convenient to combine $H^{(2)}$ and $W_{\bar z}^{(2)}$ 
into a column vector
\beq
\Phi^{(2)} ~\equiv~ \ba{c} H^{(2)} \\ 
{\frac2{g}} W_{\bar z}^{(2)} \ea.
\eeq
Note that the fluctuations transform under 
the gauge transformations as
\beq
H^{(2)} \rightarrow H^{(2)} + i \Lambda H^{(0)}, \hs{10} 
W_{\bar z}^{(2)} \rightarrow W_{\bar z}^{(2)} - \D_{\bar z} \Lambda,
\eeq
where $\Lambda$ is an arbitrary hermitian matrix of order $\p_\alpha^2$. 
In order to eliminate this unphysical degrees of freedom, 
let us impose the following gauge fixing conditions for the the fluctuations
\beq
i \D_z W_{\bar z}^{(2)} + \frac{g^2}{4} H^{(2)} H^{(0) \dagger} &=& (h.c.). 
\label{eq:gauge}
\eeq
This constraint is equivalent to the condition that 
the second order fluctuation $\Phi^{(2)}$ is orthogonal 
to the unphysical gauge zero modes $\Phi_\Lambda$ 
\beq
0 ~=~ \Big< \Phi_\Lambda \,, \, \Phi^{(2)} \Big> + (h.c.), 
\hs{10}
\Phi_\Lambda ~=~ \ba{c} i \Lambda H^{(0)} \\ - \frac{2}{g} \D_{\bar z} \Lambda \ea.
\eeq
Then, the terms containing $(H^{(2)},\,W_{\bar z}^{(2)})$ 
in the fourth order effective Lagrangian ${\cal L}_{\rm eff}^{(4)}$ 
are summarized as
\beq
- 4 \Big< \,  \Delta \Phi^{(2)} \,,\, \Delta \Phi^{(2)} \, \Big> 
+\bigg[ \Big< \, \D_\alpha \Phi^{(2)} \,,\, \p^\alpha \phi^i \Phi_i \ \Big>
+ \lambda^i \Big< \, \Phi^{(2)} \,,\, \Phi_i \, \Big> + (c.c.) \bigg],
\eeq
where we have introduced the Lagrange multiplier 
$\lambda^i$ to impose the condition that the second order fluctuation 
$\Phi^{(2)}$ is orthogonal to the physical zero modes 
$\Phi_i$, as we have done in Eq.~(\ref{eq:L_higher}) in 
order to separate the massive modes from the zero modes. 
The linearized equations of motion for $H^{(2)}$ and 
$W_{\bar z}^{(2)}$ can be written as (cf. 
${\bf H} \to \Delta^\dag \Delta$ in Eq.~(\ref{eq:massive_eq}))
\beq
4 \Delta^\dagger \Delta \Phi^{(2)} + \D_\alpha ( \p^\alpha \phi^i \Phi_i ) ~=~ \lambda^i \Phi_i. 
\label{eq:4th_eom}
\eeq
Let us first determine the Lagrange multipliers $\lambda^i $. 
By taking the inner products of the zero modes $\Phi_i^\dagger$ 
and the both hand sides of the linearized equation Eq.\,\eqref{eq:4th_eom},
we find that
\beq
\p_\alpha \p^\alpha \phi^i + \Gamma^i_{jk} \p_\alpha 
\phi^j \p^\alpha \phi^k ~=~ \lambda^i,
\label{eq:lambda}
\eeq
where we have used 
\beq
\left< \Phi_j \,,\, \D_\alpha \Phi_i \right> 
~=~ \delta_\alpha g_{i \bar j} 
~=~ g_{l \bar j} \Gamma^l_{i k} \p_\alpha \phi^k, \hs{8}
\Big< \Phi_j 
\,,\, \Delta^\dagger \Delta \Phi^{(2)} \Big> 
= \Big<\Delta \Phi_j\,,\, \Delta \Phi^{(2)} \Big> = 0.
\eeq
We find that $\lambda^i=0$ is nothing but the equation of motion for
$\phi^i$ with the second order Lagrangian Eq.\eqref{eq:2nd}. 
Then, the linearized equation Eq.\,\eqref{eq:4th_eom} becomes
\beq
\Delta^\dagger \Delta \Phi^{(2)} + \frac{1}{4} \mathbf P \Big[ \D_\alpha (\p^\alpha \phi^i \Phi_i) \Big] = 0,
\label{eq:4th_eom2}
\eeq
where $\mathbf P$ the following projection operator 
which projects out the zero modes 
\beq
\mathbf P \Phi ~\equiv~ \Phi 
- \big< \Phi_j\,,\, \Phi \big> g^{\bar j i} \Phi_i.
\eeq
It seems that in order to solve Eq.\,\eqref{eq:4th_eom2} 
in terms of $\Phi^{(2)}$, 
we need to know the explicit form of the background BPS solution . 
However, we can find the following formal expression 
for $\Delta \Phi^{(2)}$ without solving the BPS equations
\begin{eqnarray}
\Delta \Phi^{(2)} ~&=&~ {\renewcommand{\arraystretch}{2}
\frac{i}{2g} 
\ba{c} \displaystyle \frac{2}{g v} \p_\alpha \phi^i \p^\alpha \phi^j ~ S^\dagger \ \left[ \nabla_i \frac{\p}{\p \phi^j} \left( \bar \p \Omega^{-1} \Omega \right) \right] H_0^{\dagger-1} 
\\ \displaystyle
i \p_\alpha \phi^i \p^\alpha \bar \phi^j \, S^{-1} \left[ \frac{\p}{\p \bar \phi^j} \left( \Omega \frac{\p}{\p \phi^i} \Omega^{-1} \right) \right] S \ea},
\label{eq:sol_Phi2} 
\end{eqnarray}
where $\nabla_i$ is the covariant derivative on the moduli space, 
which acts on $\frac{\p}{\p \phi^j} \left( \bar \p \Omega^{-1} \Omega \right)$ 
as
\beq
\nabla_i \frac{\p}{\p \phi^j} \left( \bar \p \Omega^{-1} \Omega \right) ~ \equiv~ \left( \frac{\p}{\p \phi^i} \frac{\p}{\p \phi^j} - \Gamma^k_{ij} \frac{\p}{\p \phi^k} \right) \left( \bar \p \Omega^{-1} \Omega \right). 
\eeq
We can confirm that Eq.\,\eqref{eq:sol_Phi2} satisfies 
the linearized equation Eq.\,\eqref{eq:4th_eom2} 
by checking that the following equivalent equation is satisfied
\beq
(\Delta \Delta^\dagger) \Delta \Phi^{(2)} ~=~ - \frac{1}{4} \Delta \Big[ \D_\alpha ( \p^\alpha \phi^i \Phi_i) \Big], 
\eeq
where the operator $\Delta \Delta^\dagger$ takes the form
\begin{eqnarray}
\Delta \Delta^\dagger&=& 
\left(\begin{array}{cc}
-\D^{f}_{\bar z}\D^f_z+\frac{g^2}4 (H^{(0)\dagger}H^{(0)})_r & {\bf 0} \\ 
{\bf 0} & -\D^{a}_{z}\D^a_{\bar z}+\frac{g^2}4 (H^{(0)}H^{(0)\dagger})_r
\end{array} \right).
\end{eqnarray}
The solution $\Delta \Phi^{(2)}$ is unique 
since the operator $\Delta \Delta^\dagger$
obviously has no zero mode and hence it is invertible.   
Although it has apparent singularities due to the factor $H_0^{\dagger-1}$,
we can show that the solution Eq.\,\eqref{eq:sol_Phi2} 
is smooth everywhere 
(see Appendix \ref{sec:singularity} for the proof).

It is in general not easy to solve Eq.\,\eqref{eq:sol_Phi2} 
in terms of $\Phi^{(2)}$. 
However, the explicit form of $\Delta \Phi^{(2)}$ is sufficient 
for the purpose of determining the fourth order Lagrangian 
since it can be rewritten as
\beq
\mathcal L_{\rm eff}^{(4)} 
= \int d^2 x \, \tr \left[ - \frac{1}{2g^2} 
(F_{\alpha \beta}^{(2)})^2 \right] 
+ 4 \Big< \, \Delta \Phi^{(2)} \,,\, \Delta \Phi^{(2)} \, \Big>.
\label{eq:4thL}
\eeq
By using the first order solution \eqref{eq:Firstsol}, 
the first term in the fourth order Lagrangian \eqref{eq:4thL} 
can be calculated as 
\beq
\tr \left[ - \frac{1}{2g^2} (F_{\alpha \beta}^{(2)})^2 \right] &=& \ 
\frac{2}{g^2} \tr \left[ 
\delta_{[\alpha}^\dagger (\Omega \delta_{\beta]} \Omega^{-1}
) \delta^{\dagger [\alpha} ( \Omega \delta^{\beta]} \Omega^{-1} ) \right]. 
\label{eq:Fa2}
\eeq
Substituting \eqref{eq:Fa2} and \eqref{eq:sol_Phi2} into \eqref{eq:4thL}, 
we obtain the following form of the fourth order effective Lagrangian 
for the moduli fields 
\beq
\mathcal L_{\rm eff}^{(4)} ~=~ (A_{i j \bar k \bar l} + B_{i j \bar k \bar l}) (\p_\alpha \phi^i \p^\alpha \phi^j) \overline{(\p_\beta \phi^k \p^\beta \phi^l)} + 2 B_{i j \bar k \bar l} (\p_{[\alpha} \phi^i \p_{\beta]} \phi^j) \overline{(\p^{[\alpha} \phi^k \p^{\beta]} \phi^l)},
\label{eq:general_higher_corr}
\eeq
where the tensors $A_{i j \bar k \bar l}$ and 
$B_{i j \bar k \bar l}$ are given by
\beq
A_{i j \bar k \bar l} &=& \frac{1}{g^2} \int d^2 x \, \tr \Bigg[ \frac{4}{g^2 v^2} \left( \nabla_i \frac{\p}{\p \phi^j} (\p_{\bar z} \Omega^{-1} \Omega) \right) (H_0 H_0^\dagger)^{-1} \left( \bar \nabla_k \frac{\p}{\p \bar \phi^l} (\Omega \p_z \Omega^{-1} ) \right) \Omega \Bigg], \\
B_{i j \bar k \bar l} &=& \frac{1}{g^2} \int d^2 x \, \tr \Bigg[ \frac{\p}{\p \bar \phi^k} \left( \Omega \frac{\p}{\p \phi^i} \Omega^{-1} \right) \frac{\p}{\p \bar \phi^l} \left( \Omega \frac{\p}{\p \phi^j} \Omega^{-1} \right) \Bigg].
\eeq
%%%%%%%%%%%%%%%%%%%%%%%%%%%%%%%%%%%%%%%%%%%%%%%%%%%%%%%%%%%%%%%%%%%%%%%
\section{Correction to the single vortex effective Lagrangian}
Applying the formula Eq.\eqref{eq:4thL} to 
the single vortex solution Eq.\,\eqref{eq:single_sol}, 
we obtain the following fourth order effective Lagrangian
for a single non-Abelian vortex 
\beq
\mathcal L^{(4)}_{\rm eff} = \frac{T}{8} |\p_\alpha Z \p^\alpha Z|^2 + \frac{4\pi}{g^2} ( g^{\rm FS}_{i \bar j} \p_\alpha b^i \p_{\beta} \bar b^j ) \left[ \p^{\{\alpha} Z \, \p^{\beta\}} \bar Z - \frac{1}{2} \eta^{\alpha \beta} (\p_\gamma Z \p^\gamma \bar Z) + c ( g^{\rm FS}_{k \bar l}
\p^\alpha b^k \p^\beta \bar b^l ) \right].
\label{eq:4th}
\eeq
where $T = 2 \pi v^2$ is the vortex tension and $c$ is a constant given by
\beq
c = \frac{1}{2\pi} \int d^2 x \, \left( 1 - |z-Z|^2 e^{-\psi} \right)^2 
\sim 0.830707 \times \frac{1}{g^2 v^2} .
\eeq
\if0
The tensors $A_{ij\bar k\bar l}, B_{ij\bar k\bar l}$ must be 
invariant under the rotation $Z\to e^{i\theta}Z$ and therefore,
for instance, the tensors with a set of indices 
$\{\phi^i,\phi^j,\bar \phi^k,\bar \phi^l \}=\{Z,b,\bar b,\bar b\}$ 
automatically vanish. 
The first line in the r.h.s.\,of Eq.\eqref{eq:4th} can be derived
without using the explicit form Eq.\eqref{eq:single_sol}.
Translational symmetry tells us that the derivative 
$\frac{\p}{\p Z}$ can be
rewritten to $-\p_z$ on $\Omega$, and then, 
the master equation Eq.\eqref{eq:master} can reduce the number of the
derivative $\p_z, \p_{\bar z}$. After 
some terms are canceled out between 
$A_{ij\bar k\bar l}$ and $ B_{ij\bar k\bar l}$, finally we find that 
remaining terms can be integrated by using the formula for the metric 
Eq.\eqref{eq:metric} and so on.
To get the last term of Eq.\eqref{eq:4th}, we used the following.
The tensor $A_{bb\bar b \bar b}$ vanish since 
non-vanishing elements of $\frac{\p}{\p \bar b^i} (\Omega\p_z \Omega^{-1})$ 
is proportional to the metric $g^{\rm FS}_{i\bar j}$ which is
covariantly constant. The tensor $B_{bb\bar b \bar b}$ is calculated 
by using $g_{i\bar j}^{\rm FS}=\p_{b^i}\p_{\bar b^j}\ln X$ as
\begin{eqnarray}
B_{b^ib^j\bar b^k\bar b^l}= \frac{2\pi}{g^2}c \left(g^{\rm FS}_{i\bar k}g^{\rm FS}_{j\bar l}
+g^{\rm FS}_{i\bar l}g^{\rm FS}_{j\bar k}\right),
\end{eqnarray}
and therefore this tensor is symmetric in indices $i,j$ ($\bar k,\bar l$).
\fi
In summary, we have obtained the following derivative expansion of 
the effective Lagrangian for a single non-Abelian vortex
\beq
\mathcal L_{\rm eff} \hs{-2} &=& - \ T \ \left[ 1 - \frac{1}{2} \p_\alpha Z \p^\alpha \bar Z - \frac{1}{8} |\p_\alpha Z \p^\alpha Z|^2 \right] \phantom{\Bigg[} \notag \\
&& + \frac{4\pi}{g^2} \left[ \eta^{\alpha \beta} \Big( 1 - \frac{1}{2} \p_\beta Z \, \p^\beta \bar Z \Big) + \p^{\{\alpha} Z \, \p^{\beta\}} \bar Z \right] \left( g^{\rm FS}_{i \bar j} \p_\alpha b^i \p_{\beta} \bar b^j \right) \phantom{\Bigg[} \notag \\
&& + \frac{4\pi c}{g^2} \left( g^{\rm FS}_{i \bar j}
\p_\alpha b^i \p_\beta \bar b^j \right) \left( g^{\rm FS}_{k \bar l}
\p^\alpha b^k \p^\beta \bar b^l \right) + \mathcal O(\p_\alpha^6). \phantom{\Bigg[} \label{eq:sum}
\eeq
%%%%%%%%%%%%%%%%%%%%%%%%%%%%%%%%%%%%%

\subsection{Comparison with other models}\label{sec:com}

Let us compare our result, Eq.\,(\ref{eq:sum}), 
of the four derivative terms in the effective action 
for a single non-Abelian vortex with those in different context. 
First we compare the translation modes $Z$ with the Nambu-Goto action, 
and second we compare the orientational modes $b^i$ with 
the Faddeev-Skyrme model. 

\paragraph{Translational zero modes and mixed terms \\}
First, by setting the orientational moduli $b^i$ to zero 
in Eq.\,(\ref{eq:sum}), we have the effective action 
for the translational modes $Z$ as
\beq
S_{Z} = - 2 \pi v^2 \int d^{d-1} x 
\left( 1 - {1\over 2} \p_\alpha Z \p^\alpha \bar Z  
- {1\over 8} |\p_\alpha Z \p^\alpha Z|^2 \right), 
\eeq
which is precisely the one of an Abelian (ANO) vortex.
At this order, this effective Lagrangian coincides with 
the Nambu-Goto action \cite{Nambu:1974zg} given by 
\beq
S_{\rm NG} = - T \int d^{d-1} x \, \sqrt{- \det (- \gamma_{\alpha \beta})},
\eeq
where $\gamma_{\alpha \beta}$ 
is the induced metric on the world-volume given by 
\beq
\gamma_{\alpha \beta} ~=~ \partial_{\alpha} X^\mu \partial_{\beta} X^\nu \eta_{\mu \nu} 
~=~ \eta_{\alpha \beta} - \frac{1}{2} \left( \p_\alpha Z \p_\beta \bar Z + \p_\alpha \bar Z \p_\beta Z \right).
\eeq
In Eq.\,(\ref{eq:sum}),  
there are the four derivative terms containing 
both the translational zero modes $Z$ and 
the orientational zero modes $b^i$.
These are precisely the terms appearing in the derivative expansion 
of the following action describing the $\C P^{N-1}$ sigma model 
on the vortex world-volume
\beq
S_{\rm eff} = \int d^{d-1} x \, \sqrt{- \det (- \gamma_{\alpha \beta})} \left[ - T + \frac{4\pi}{g^2} \gamma^{\alpha \beta} (g^{\rm FS}_{i \bar j} \p_\alpha b^i \p_\beta \bar b^j) + \mathcal O(\p_a^4) \right].
\eeq

\paragraph{Orientational zero modes \\}
Next, we consider the orientational zero modes in the internal space.
For simplicity, let us restrict ourselves to the case of $N=2$
in which the internal moduli space is ${\mathbb C}P^{1}$. 
By setting $Z=0$ in Eq.\,(\ref{eq:sum}), 
we obtain the following effective Lagrangian of the orientational zero modes 
with the higher derivative correction
\beq
\mathcal L_{{\mathbb C}P^1} = \frac{4\pi}{g^2} \left[ \frac{\p_\alpha b \, \p^\alpha \bar b}{(1+|b|^2)^2}
+ c \frac{|\p_\alpha b \, \p^\alpha b|^2}{(1+|b|^2)^4} \right]. 
\label{eq:CPN-1+four}
\eeq
Now let us compare this with the Skyrme-Faddeev model which is 
a ${\mathbb C}P^{1}$ model with a four-derivative term 
\cite{Faddeev:1996zj}. 
To this end, we formulate the ${\mathbb C}P^1$ model 
(at the leading order) by two complex fields 
$h =(h^1,h^2)$ charged under $U(1)$ gauge symmetry.
By introducing auxiliary gauge field $a_{\alpha}$ and 
scalar field $\lambda$ as Lagrange multipliers, 
it can be written as
\beq
{\cal L} ~=~ (\p_{\alpha} + i a_{\alpha}) h \, (\p_{\alpha} - i a_{\alpha}) h^\dagger - \lambda \left(h h^\dagger - {4 \pi \over g^2} \right)
 \label{eq:CP^1-Kahler-quot}
\eeq
The variation of $\lambda$ gives the constraint 
$hh^\dagger = {4 \pi \over g^2}$, 
which can be solved by $h = \sqrt{{(g^2/4\pi) \over {1+|b|^2}}}(1,b)$.
Then the variation of $a_{\alpha}$ gives 
\beq
&& a_{\alpha} 
 = \frac{i}{2}{\bar b \partial_{\alpha} b - b \partial_{\alpha} \bar b 
 \over (1 + |b|^2)} .\label{eq:Amu}
\eeq
Substituting these back into the original Lagrangian 
(\ref{eq:CP^1-Kahler-quot}), the ${\mathbb C}P^1$ model (at the leading order) is recovered. 

The field strength of the gauge field (\ref{eq:Amu}) is
\beq
f_{\alpha \beta} ~=~ \partial_{\alpha} a_{\beta} - \partial_{\beta} a_{\alpha} ~=~-i{\partial_{\alpha} b \, \partial_{\beta} \bar b - \partial_{\beta} b \, \partial_{\alpha} \bar b \over (1+|b|^2)^2 }. \label{eq:Fmunu}
\eeq
Then the Faddeev-Skyrme term \cite{Faddeev:1996zj} 
can be written as the field strength squared as\footnote{
This term is also called the baby Skyrme term in 
$d=2+1$ \cite{Piette:1994ug}. 
}
\beq
f_{\alpha \beta} f^{\alpha \beta} 
~=~ 2 {(\partial_{\alpha} \bar b \, \partial^{\alpha} b)^2 - |\partial_{\alpha} b \, \partial^{\alpha} b|^2  \over (1+|b|^2)^4} .
\eeq
This term is quadratic in the time derivative and 
does not coincide with Eq.\,(\ref{eq:CPN-1+four}). 
The other term containing four time-derivatives appears in 
an ${\cal N}=1$ supersymmetric extension 
of the Skyrme-Faddeev term \cite{Freyhult:2003zb}:
\footnote{
This term also arises when one constructs the Faddeev-Skyrme-like model 
as the low-energy effective theory of pure $SU(2)$ Yang-Mills theory
\cite{Gies:2001hk}, and hence is called the Gies term in that context.
}
\beq
{(\partial_{\alpha} b \, \partial^{\alpha} \bar b)^2 \over (1+|b|^2)^4}.
\eeq
As shown in Appendix \ref{appendix:higher_CP}, 
this higher derivative term can be obtained 
by adding a higher derivative term to the original Lagrangian 
\eqref{eq:CP^1-Kahler-quot}. 
In general the fourth order terms are summarized as
\beq
{\cal L}^{(4)}_{\rm general} ~=~ c_1 f_{\alpha \beta} f^{\alpha \beta} + c_2 {(\partial_{\alpha} b \, \partial^{\alpha} \bar b)^2 \over (1+|b|^2)^4}
~=~ { (2c_1+c_2)(\partial_{\alpha} \bar b \, \partial^{\alpha} b)^2 
- 2 c_1 |\partial_{\alpha} b \, \partial^{\alpha} b|^2 \over (1+|b|^2)^4} .
\label{eq:four-deriv-general}
\eeq
This reduces to the four derivative term in Eq.\,(\ref{eq:CPN-1+four}) 
when
\beq 
c_2 = - 2 c_1 = c.
\eeq
The condition $c_2 = -2c_1$ is precisely the condition 
for ${\cal N}=1$ supersymmetry \cite{Freyhult:2003zb}.
This must be the case because
there remains ${\cal N}=1$ supersymmetry (four supercharges) 
in the vortex effective theory, 
because vortices are 1/2 BPS states in supersymmetric 
theories with eight supercharges.

%%%%%%%%%%%%%%%%%%%%%%%%%%%%%%%%%%%%%%%%%%%%
\subsection{Instantons trapped inside a non-Abelian vortex}\label{sec:inst}
In this section, we consider $d=4+1$ dimensions where 
vortices are membranes with 2+1 dimensional world-volume.
Besides the vortices, there also exist Yang-Mills instantons 
which are particle-like solitons in $d=4+1$ dimensions.
But they cannot exist stably in the Higgs phase which we 
are considering. 
Instead, they can stably exist inside the world-volume of 
a non-Abelian vortex \cite{Eto:2004rz,Fujimori:2008ee}.
Instanton-vortex composite configurations are 1/4 BPS states in supersymmetric gauge theories with 8 supercharges. 
In the supersymmetric ${\mathbb C}P^{N-1}$ model with 4 supercharges, 
at the leading order of the 1/2 BPS vortex effective theory, 
instantons can be regarded as 1/2 BPS sigma model lumps 
\cite{Eto:2004rz} which are point-like solitons in $d=2+1$ 
dimensions. 
Here we show that instantons (= lumps) are not modified even 
if we include derivative corrections found in the previous sections.

For simplicity, we set $Z=0$ and consider 
the case of $N=2$ (the $\C P^1$ sigma model)
with the Lagrangian (\ref{eq:CPN-1+four}) for the orientational modes.
In $d=4+1$ dimensions, the vortex is a membrane 
which has coordinates $(x^0=t,x^1,x^2)$. 
We parametrize the two spatial coordinates by 
complex variables $w=x^1 + ix^2$ and $\bar w = x^1 - i x^2$.
We discuss only static configurations in this model. 

First we recall the lump solution in the model without 
the four derivative correction. 
The energy density can be written as
\beq
E_{{\mathbb C}P^1}^{(2)} 
&=& \frac{4\pi}{g^2} \int d^2 x \sum_{\alpha=1,2} \frac{\p_\alpha b \, \p^\alpha \bar b}{(1+|b|^2)^2}
~=~ \frac{8\pi}{g^2} \int d^2 x \frac{ 2 |\p_{\bar w} b|^2 + (|\p_w b|^2 - |\p_{\bar w} b|^2)}{(1+|b|^2)^2}.
 \label{eq:E-lumps}
\eeq
The first term is positive semi-definite $|\p_{\bar w} b|^2 \geq 0$
and the second term can be rewritten as
\beq
I ~=~ \frac{4\pi i}{g^2} \int \frac{\p_w b \, \p_{\bar w} \bar b - \p_{\bar w} b \p_w \bar b}{(1+|b|^2)^2} \, dw \wedge d\bar w  
~=~ \frac{4\pi i}{g^2} \int \frac{d b \wedge d \bar b}{(1+|b|^2)^2} ~=~ \frac{8\pi^2}{g^2} k,
\eeq
where the integer $k$ is the degree of the map from 
the vortex world-volume to the target space $\C P^1$. 
This is the topological charge for lumps: 
$\pi_2({\mathbb C}P^1) = {\mathbb Z}$.
The energy (\ref{eq:E-lumps}) is bound from below 
by this topological charge $I$, 
and the Bogomol'nyi bound is saturated 
by the BPS equation for lumps, given by
\beq
\p_{\bar{w}} b = 0.
\label{eq:BPS-lump}
\eeq
The BPS solutions $b(w)$ are holomorphic in $w$
and satisfy the static equation of motion. 
If we fix the boundary condition as 
$b \rightarrow \infty~(w \rightarrow \infty)$, 
the solutions with $I = \frac{8\pi^2}{g^2} k$ 
are given by $k$-th order rational maps
\beq
b(w) ~=~ \frac{a_0 w^k + a_1 w^{k-1} + \cdots + a_k}{\tilde a_0 w^{k-1} + \tilde a_1 w^{k-2} + \cdots + \tilde a_{k-1} }.
\label{eq:holo_map}
\eeq
This is the general solutions for the 1/2 BPS sigma model lumps. 

Next let us consider the effect of the four derivative term 
\beq
E_{{\mathbb C}P^1}^{(4)} &=& 
-\frac{64\pi c}{g^2} \int d^2 x \, \frac{|\p_w b \, \p_{\bar w} b|^2}{(1+|b|^2)^4}.
\eeq
The total energy can be rewritten as
\beq
E_{\C P^1}^{(2)+(4)} ~=~ \frac{16 \pi}{g^2} \int d^2 x \left[ 1 - 4 c \frac{|\p_w b|^2}{(1+|b|^2)^2} \right] \frac{|\p_{\bar w} b|^2}{(1+|b|^2)^2} ~+~ \frac{8 \pi^2}{g^2} k.
\eeq
Again, the first term is positive semi-definite 
as long as the correction term is sufficiently small and 
vanishes if the BPS equation (\ref{eq:BPS-lump}) is satisfied.
We thus have found that there is no contribution 
from the four derivative term to the BPS configurations 
and consequently the lump solutions are not modified. 
The energy of lumps $\frac{8\pi^2}{g^2} k$ agrees 
with that of instantons in the Higgs phase, 
which can be found by rewriting the energy of the original bulk theory 
to the Bogomol'nyi form for 1/4 BPS configurations.
Therefore, it is natural to conjecture that 
all higher order terms vanish for 1/2 BPS lump configurations.

By considering 1/2 BPS lumps on the ${\mathbb C}P^{N-1}$ model, 
we have obtained 1/4 BPS composite states of instantons inside a vortex.
On the other hand, there exist another 1/4 BPS composite states 
of intersecting vortex-membranes \cite{Eto:2004rz,Fujimori:2008ee}. 
Vortices of one kind have codimensions in $z$-plane and extend to $w$-plane, 
and those of another kind have codimensions 
in $w$-plane and extend to $z$-plane. 
This intersecting vortices can be constructed by 
considering holomorphic maps of the translational moduli $Z(w)$, 
instead of the orientational moduli $b^i(w)$ considered in this section.
Now let us show that, in general, 
the fourth derivative terms do not modify the 1/2 BPS states. 
The BPS equation can be found by rewriting the second order energy as
\beq
E_{\rm eff}^{(2)} = \int d^2 x \, 4 g_{i \bar j} \p_{\bar w} \phi^i \p_w \bar \phi^j + \int i g_{i \bar j} d \phi^i \wedge d \bar \phi^j.
\eeq
The second term gives the area of a two cycle of the moduli space 
on which the map $\phi^i$ wraps. 
The first term is positive semi-definite 
and vanishes for any holomorphic maps satisfying
\beq
\p_{\bar w} \phi^i ~=~ 0.
\eeq
On the other hand, 
it follows from the general form of 
the higher derivative corrections 
Eq.\,\eqref{eq:general_higher_corr} 
that the fourth order energy vanishes for the holomorphic maps
\beq
E_{\rm eff}^{(4)} ~\propto~ 
\p_{\bar w} \phi^i \p_w \phi^j \p_w \bar \phi^k \p_{\bar w} \bar \phi^l ~=~ 0.
\eeq
Therefore, the higher derivative corrections do not modify 
the 1/2 BPS configurations. 

There is a possibility that 
the holomorphic maps are unstable if the fourth order energy is negative.
However, this would be an artifact of the truncation at the fourth order.
For example, the translational part of the energy can be rewritten as
\beq
E_{\rm eff} ~=~ T \int d^2 x \, 2 (1 - |\p_w Z|^2) |\p_{\bar w} Z|^2 + T \int \frac{i}{2} \left( d w \wedge d \bar w + d Z \wedge d \bar Z \right). 
\eeq
The second term is a topological term and 
the first term vanishes for a holomorphic map $Z(w)$ 
but it is not positive semi-definite. 
Although a holomorphic map $Z(w)$ appears to be an unstable solution 
if $|\p_w Z|^2 > 1$, such a case is beyond the validity of our approximation. 
We can show that all the holomorphic maps are stable BPS solutions 
if we consider the full order action, i.e. the Nambu-Goto action, 
whose energy can be rewritten as
\beq
E_{\rm NG} &=& T \int d^2 x \left[\sqrt{(1 + |\p_{\bar w} Z|^2 - |\p_w Z|^2)^2 + 4 |\p_{\bar w} Z|^2} - (1 + |\p_{\bar w} Z|^2 - |\p_w Z|^2) \right] \notag \\
&+& T \int \frac{i}{2} \left( d w \wedge d \bar w + d Z \wedge d \bar Z \right). 
\eeq
The first term is positive semi-definite and can be expanded as
\beq
T \int d^2 x \, 2 \Big[ 1 - |\p_w Z|^2 + |\p_w Z|^2 (|\p_w Z|^2 + |\p_{\bar w} Z|^2) + \cdots \Big] |\p_{\bar w} Z|^2.
\eeq
As this example shows, we should take into account 
the full order corrections 
to prove the stability of the holomorphic maps. 

\subsection{Higher derivative terms and massive modes}\label{sec:massive mode}
\begin{figure}[h]
\centering
\includegraphics[width=40mm]{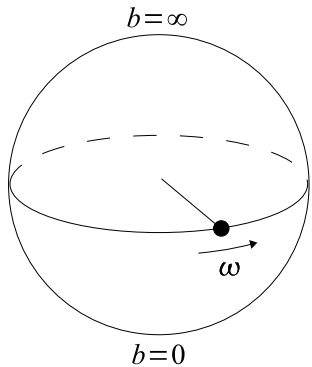}
\caption{{\footnotesize The rotation of the orientation}}
\label{fig:rotation}
\end{figure}

\begin{figure}[h]
\begin{center}
\begin{tabular}{cc}
\includegraphics[width=60mm]{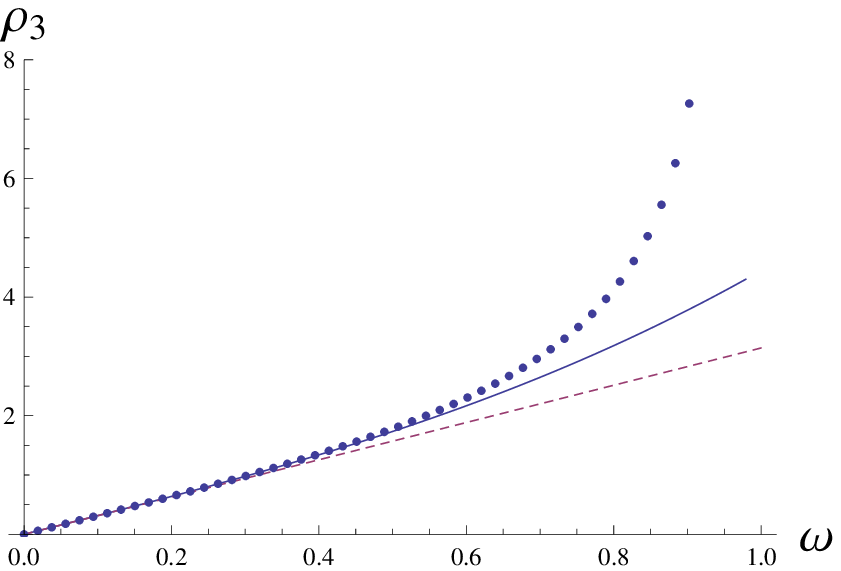} \hs{8} & \hs{8}
\includegraphics[width=63mm]{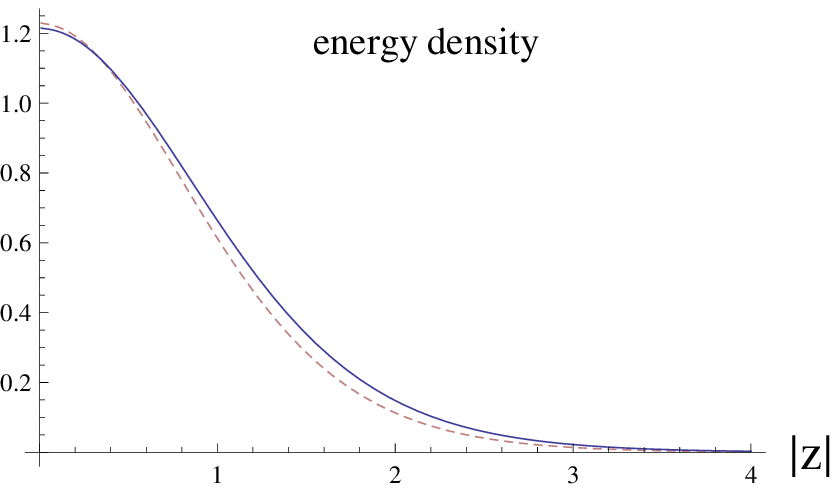} \\ 
(a) The relation between $\rho_3$ and $\omega$ \hs{8} & \hs{8} 
(b) The shift of the vortex size
\end{tabular}
\end{center}
\caption{{\footnotesize (a) The relation between the angular velocity $\omega$ 
and the conserved charge $\rho_3$ for $g=v=1$. 
Compared to the second order relation 
$\rho_3 = \frac{\pi}{g^2} \omega$ (dashed line), 
the fourth order relation 
$\rho_3=\frac{\pi}{g^2}(1 + \frac{c}{2} \omega^2) \omega$ (solid line) 
shows the better agreement with the numerical result (dots) 
obtained by solving the full equations of motion 
Eqs.\,\eqref{eq:eom_full1} and \eqref{eq:eom_full2}.
(b) The energy density distributions in the transverse plane 
($z$-plane) for $g=v=1$.
The size of the excited vortex with $\omega = 0.5$ (solid line) 
is slightly larger than that of the static vortex with $\omega = 0$ 
(dashed line).}}
\label{fig:charge}
\end{figure}

To see the physical meaning of the higher derivative corrections 
for the internal orientation, let us consider the equation 
of motion with higher order corrections in the $N=2$ case. 
Assuming that the orientational zero mode $b$ 
is independent of the spatial world-volume coordinates, 
we can show that the following configuration satisfies the 
equation of motion even in the presence of the higher 
derivative terms : 
\beq
b ~=~ \exp \left( i \omega t \right).
\eeq
This solution corresponds 
to an excited state of the non-Abelian vortex 
whose orientation is rotating along
the equator of $\C P^1$ (Fig.\,\ref{fig:rotation}). 
The rotation of the orientation induces
one of three components of the $SU(2)_{\rm C+F}$ conserved charges, 
which is given by
\beq
\rho_3 ~=~ \frac{i}{2} \left( \frac{\p \mathcal L}{\p (\p_t b)} b - \frac{\p \mathcal L}{\p (\p_t \bar b)} \bar b\right) ~=~ \frac{\pi}{g^2} \left( 1 + \frac{1}{2} c \, \omega^2 \right) \omega.
\eeq
The relation between the angular velocity $\omega$ 
and the conserved charge $\rho_3$ (shown in Fig.\,\ref{fig:charge}-(a)) 
is modified by the higher derivative term, 
as in the case of the angular velocity 
and the angular momentum of the particle (see Eq.\,\eqref{eq:angular}). 

By numerically solving the full equations of motion 
Eqs.\,\eqref{eq:eom_full1} and \eqref{eq:eom_full2}, 
we can see that the size of the excited vortex is slightly larger 
than that of the static configuration (see Fig.\,\ref{fig:charge}-(b)). 
This is analogous to the case of the particle
discussed in Sec.\,\ref{sec:prelim}: 
the higher derivative term corresponds to 
the correction from the massive mode (vortex size) 
which is slightly shifted by the conserved charge. 

%%%%%%%%%%%%%%%%%%%%%%%%%%%%%%%%%%%%%%%%%%%%%%%%%%%%%%%%%%%%%%%%%%%%
\section{Mass deformations}\label{sec:mass}

\subsection{Higher order corrections from mass terms}
In this section, we consider higher order corrections 
from supersymmetric mass deformations of the original theory.  
Without breaking the supersymmetry, we can deform the original model  
by adding the following terms to the Lagrangian Eq.\,\eqref{eq:action} 
\beq
\mathcal L_{{\rm adjoint} + {\rm mass}} 
~=~ \tr \left[ \frac{1}{g^2} \D_\mu \Sigma_I \D^\mu \Sigma_I 
+ \frac{1}{2g^2} [ \Sigma_I , \Sigma_J ]^2 - |\Sigma_I H - H M_I |^2 \right],
\label{eq:mass}
\eeq
where $\Sigma_I$ ($I=1,\cdots,n$) are the real adjoint scalar 
fields in the vector multiplets and $M_I$ are mass matrices 
in the Cartan subalgebra of $SU(N)_{\rm F}$
\beq
M_I ~\equiv~ \mathbf m_I \cdot \mathbf H 
~=~ {\rm diag} (m_{I,0},\,m_{I,1},\,\cdots,\,m_{I,N-1}).
\eeq
If all the masses are non-degenerate, 
the $SU(N)_{\rm F}$ flavor symmetry is maximally broken to 
the Cartan subalgebra of $SU(N)_{\rm F}$. 
These mass deformations can be obtained by dimensional reductions 
from the six dimensional model on $\mathbb R^{6-n} \times (S^1)^n$ 
with the following twisted boundary conditions around periodic dimensions
\beq
H(x^\mu, \theta_I + 2 \pi R_I) = H (x^\mu, \theta_I) e^{2 \pi i R_I M_I}.
\eeq
Ignoring the infinite tower of the Kaluza-Klein modes, 
we obtain the mass deformation Eq.\,\eqref{eq:mass} with 
the following identification
\beq
\Sigma_I(x^\mu) = - W_{\theta_I}(x^\mu), \hs{10} 
H(x^\mu) e^{ i \theta_I M_I } = H(x^\mu, \theta_I).
\eeq
The mass terms do not change the VEV of $H$ (given in Eq.\,\eqref{eq:H_VEV}),
while they induce those of the adjoint scalars 
\beq
\langle \Sigma_I \rangle = M_I.
\eeq
In this mass deformed model, 
the color-flavor global symmetry $SU(N)_{\rm C+F}$ is 
explicitly broken to the Cartan subgroup $U(1)^{N-1}$. 
Hence the orientational moduli, which were the Nambu-Goldstone zero modes 
of $SU(N)_{\rm C+F}$, are lifted by a potential 
induced by the mass terms. 
The potential on the moduli space can be calculated 
by finding the minimum energy configuration 
and evaluating the energy for each values of the moduli parameters.
This can be done perturbatively with respect to the masses $m_{I,A}$. 
As a zeroth order configuration, 
we consider the vortex solution Eq.\,\eqref{eq:BPSsol}
satisfying the BPS equations without the mass deformations. 
Then we can determine the corrections to $\Sigma_I$ 
by solving their equations of motion
\beq
\frac{2}{g^2} \Big( \D_\mu \D^\mu \Sigma_I + \big[ \Sigma_J \big[ \Sigma_J, \Sigma_I \big] \big] \Big) = (\Sigma_I H - H M_I ) H^\dagger + H (H^\dagger \Sigma_I - M_I H^\dagger ). 
\eeq
Due to the fact that $\Sigma_I$ originate from
the higher dimensional gauge fields $W_{\theta_I}$, 
we can solve the equations of motion 
in the similar way to the gauge fields $W_\alpha$
\beq
\Sigma_I^{(1)} = M + i ( \delta_I S^\dagger S^{\dagger-1} - S^{-1} \delta_I^\dagger S ) + \mathcal O(m^3, m \p_\alpha^2) ,
\eeq
where we have assumed that the derivatives $\p_\alpha$ 
and the masses $m$ are of the same order
\beq
m ~\sim~ \p_\alpha.
\eeq
The differential operators $\delta_I$ and $\delta_I^\dagger$ are defined by
\beq
\delta_I = k_I^i \frac{\p}{\p \phi^i}, \hs{10}
\delta_I^\dagger = \bar k_I^i \frac{\p}{\p \bar \phi^i},
\eeq
where $k_I^i$ are holomorphic Killing vectors on the moduli space 
which are the following linear combinations of the Killing vectors of the 
unbroken symmetry $U(1)^{N-1}$
\beq
k_I^i ~\equiv~ \mathbf m_I \cdot \boldsymbol \xi^i. 
\eeq
For example, the holomorphic Killing vectors on the moduli space of 
the single vortex solutions are given by
\beq
k_I^i ~=~ i ( m_{I,i} - m_{I,0} ) b^i~~(\mbox{no sum over $i$}).
\eeq

Inserting the solution $\Sigma_I^{(1)}$ 
into the deformation terms Eq.\,\eqref{eq:mass}, 
we obtain the effective potential of the form
\beq
V_{\rm eff}^{(2)} = g_{i \bar j} \, k_I^i \bar k_I^j.
\label{eq:potential_2nd}
\eeq
This is the sum of the squared norm of the Killing vectors $k_I^i$
evaluated with respect to the moduli space metric $g_{i \bar j}$ 
given in Eq.\,\eqref{eq:metric}. 

The potential of this form can also be obtained by dimensional reductions 
from the four dimensional vortex world-volume action 
(the effective theory of the vortex in six spacetime dimensions) 
\beq
S_{\rm eff}^{4d} ~=~ \int d^4 x \, g_{i \bar j} 
\Big[ \p_\alpha \phi^i \p^\alpha \bar \phi^j 
- \p_{\theta_I} \phi^i \p_{\theta_I} \bar \phi^j \Big].
\eeq
As in the case of the bulk theory, we impose the twisted boundary condition 
for the moduli fields
\beq
\phi^i(x^\alpha, \theta_I + 2 \pi R_I) = \phi^i(x^\alpha, \theta_I) 
e^{2 \pi R_I k_I^i}.
\eeq
Keeping only the lowest modes and 
evaluating the kinetic terms in the effective action,
we obtain the effective action with the potential Eq.\,\eqref{eq:potential_2nd}
\beq
S_{\rm eff}^{(4-n)d} ~=~ \int d^{4-n} x \, g_{i \bar j} \Big[ \p_\alpha \phi^i \p^\alpha \bar \phi^j - k_I^i \bar k_I^j \Big]. 
\eeq
Therefore, the procedure of the dimensional reductions 
and the calculation of the effective action are commutative. 

The higher order corrections to 
$H$ and $W_{\bar z}$ can also be determined from
the equations of motion with the mass deformations. 
They can also be solved in a similar way to 
the case without mass deformation as
\beq
\Delta \Phi^{(2)} ~=~ \frac{i}{2g} 
{\renewcommand{\arraystretch}{2}
\ba{c} \displaystyle 
\frac{2}{g v} \, \left[ \p_\alpha \phi^j \p^\alpha \phi^k - k_I^j k_I^k \right] ~ \, S^\dagger \ \left[\nabla_j \frac{\p}{\p \phi^k} \left( \bar \p \Omega^{-1} \Omega \right) \right] H_0^{\dagger-1} 
\\ \displaystyle
\ i \, \left[ \p_\alpha \phi^j \p^\alpha \bar \phi^k - k_I^j \bar k_I^k \right] \, S^{-1} \left[ \frac{\p}{\p \bar \phi^k} \left( \Omega \frac{\p}{\p \phi^j} \Omega^{-1} \right) \right] S \ea}. 
\eeq
Then the higher order corrections are obtained by 
substituting the solution into 
\beq
\mathcal L^{(4)}_{\rm eff} = \int d^2 x \, \tr \left[ - \frac{1}{2g^2} (F_{\alpha \beta}^{(2)})^2 + \frac{1}{g^2} (\D_\alpha \Sigma_I^{(1)})^2 + \frac{1}{2g^2} [\Sigma_I^{(1)}, \Sigma_J^{(1)}]^2 \right] + 4 \Big< \Delta \Phi^{(2)} \,,\, \Delta \Phi^{(2)} \Big>.
\eeq
In the case of the effective Lagrangian of a single vortex, 
the induced terms in take the form
\beq
\mathcal L_{\rm eff \, mass}^{(2) + (4)} &=& - \frac{4\pi}{g^2} g_{i \bar
j}^{\rm FS} k_I^i \bar k_I^j \Big( 1 - \frac{1}{2} \p_\alpha Z \p^\alpha
\bar Z \Big)\notag \\
&& + \frac{4\pi c}{g^2} \Big[ (g^{\rm FS}_{i \bar j} k_I^i \p_\alpha \bar b^j) 
\, (g^{\rm FS}_{k \bar l} k_I^k \p^\alpha \bar b^l) +{\rm c.c.} \Big] \notag \\
&& + \frac{4\pi c}{g^2}  (g^{\rm FS}_{i \bar j} k_I^i \bar k_J^j) 
(g^{\rm FS}_{k \bar l} k_I^k \bar k_J^l) .
\eeq
These corrections terms can also be obtained by 
using the dimensional reductions from the four-dimensional world-volume 
with the higher derivative corrections. 

\subsection{Kink monopoles}

In this section we consider a single 1/2 BPS non-Abelian 
vortex-string in four spacetime dimensions and its low-energy effective 
theory in two spacetime dimensions. 
Besides the vortex, an another important topological 
soliton, 1/2 BPS monopole, arises in four dimensions.
However, since the gauge symmetry is completely broken 
in the Higgs phase, the monopole cannot stay alone and must 
be accompanied with the vortex-string.
If the vortex-strings are attached to the monopoles 
from the both left and right hand sides, 
the composite vortex-monopole state is indeed 
a stable BPS state preserving 1/4 supersymmetry 
(two supercharges) \cite{Tong:2003pz}.
Such confined monopoles can also be viewed as 
kinks in the two-dimensional vortex effective theory.
For example, let us consider $N=2$ case with only one mass term 
\beq
M ~=~ \frac{1}{2}(m,-m).
\eeq
The orientational part of the second order effective Lagrangian 
of a single vortex-string in the mass deformed $U(2)$ theory takes the form 
\beq
\mathcal L_{\rm eff}^{(2)} ~=~ \frac{4\pi}{g^2} 
\frac{\p_\alpha b \, \p^\alpha \bar b - m^2 |b|^2}{(1+|b|^2)^2},
\qquad (\alpha = 0,1).
\label{eq:eff_2nd}
\eeq
In this sigma model, the orientational moduli are lifted 
by the potential and there are discrete vacua at $b=0,\infty$. 
In the $(1+1)$-dimensional vortex world-sheet, we can consider 
a BPS kink interpolating between the discrete vacua 
\beq
E_{\rm eff}^{(2)} = \int dx \left[ \frac{4\pi}{g^2} 
\frac{|\p_x b - m b|^2}{(1+|b|^2)^2} + \p_x \sigma \right],
\eeq
where $\sigma$ is the moment map of the unbroken $U(1)$ 
symmetry defined by 
\beq
\sigma = - \frac{2\pi m}{g^2} \frac{1-|b|^2}{1+|b|^2}.
\eeq
The BPS equation can be easily solved as
\beq
\p_x b = m b, ~~~~~\longrightarrow~~~~~
b(x) = \exp \left[ m(x-x_0)+i\theta \right], \label{eq:kink_sol}
\eeq
where $x_0$ and $\theta$ are kink position and phase moduli, 
respectively. 
The mass of the kink is 
\beq
M_{\rm BPS} ~=~ \sigma(x \rightarrow \infty) 
- \sigma(x \rightarrow - \infty) ~=~ \frac{4\pi m}{g^2}. 
\label{eq:kink_mass}
\eeq
As expected, this BPS mass of the kink 
coincides with that of the monopole in $(3+1)$ dimensions. 

To see the effect of the higher order corrections to the BPS kinks, 
let us take into account the forth order corrections
in the effective theory
\beq
\mathcal L_{\rm eff}^{(2+4)} = \frac{4\pi}{g^2} 
\left[ \frac{\p_\alpha b \, \p^\alpha \bar b - m^2 |b|^2}{(1+|b|^2)^2} 
+ c \frac{|\p_\alpha b \, \p^\alpha b + m^2 b^2 |^2}{(1+|b|^2)^4} \right].
\label{eq:eff_lag_fourth}
\eeq
The energy of static configurations can be rewritten as 
\beq
E_{\rm eff}^{(2+4)} =  \int dx \left[ \frac{4\pi}{g^2} 
\frac{|\p_x b - m b|^2}{(1+|b|^2)^2} 
\left( 1 - c \frac{|\p_x b + m b|^2}{(1+|b|^2)^2} \right) 
+ \p_x \sigma \right].
\eeq
Therefore, the BPS kink solution and its mass 
Eqs.\,\eqref{eq:kink_sol} and \eqref{eq:kink_mass} 
are not modified by the higher order corrections. 
Thus, we again encountered the case where the topological 
soliton in the vortex effective theory is exact at the second order. 
This is consistent with the observation that the kink 
on the vortex corresponds to the monopoles attached by 
the vortices in the four dimensional full theory. 

%%%%%%%%%%%%%%%%%%%%%%%%%%%%%%%%%%%%%
\subsection{Q-solitons}

Let us again consider the effective theory of 
the single non-Abelian vortex in $4+1$ dimensions.
In the case of $N=2$, 
the effective theory is ($2+1$)-dimensional 
massive $\mathbb{C}P^1$ sigma model.
The effective Lagrangian at the second order 
is the same as Eq.~(\ref{eq:eff_2nd}) with 
$\alpha = 0,1,2$. As is well known, from the Derrick's theorem, 
there are no stable lump solitons: 
the potential makes them collapse. 
However, one can still construct stable solitons by adding 
Noether charges to the lumps. 
It is the so-called Q-lumps~\cite{Leese:1991hr,Abraham:1991ki}.
The Q-lump solutions have the same form as the $\mathbb{C}P^1$ 
but their phases are time-dependent. 
To see this, let us see the energy 
\beq
E^{(2)} &=& \frac{4\pi}{g^2}\int d^2x\ \frac{ |\dot b|^2 + |\p_i b|^2 - m^2 |b|^2}{(1+|b|^2)^2} \nonumber\\
&=& \frac{4\pi}{g^2}\int d^2x\ \frac{ |\dot b \mp i m b|^2 + |\p_{\bar w} b|^2/2}{(1+|b|^2)^2} + I + Q,
\eeq
with $I$ and $Q$ being the topological charge and the Noether charge associated with the $U(1)$ global symmetry
\beq
I &=& \frac{8\pi i}{g^2}\int \frac{db \wedge d\bar b}{(1+|b|^2)^2} = \frac{8\pi^2k}{g^2},\\
Q &=& \pm \frac{4\pi m}{g^2}\int d^2x\ \frac{i ( b \dot b^* - \dot b b^*)}{(1+|b|^2)^2}.
\eeq
The BPS equations are of the form
\beq
\dot b = \pm i m b,\quad \p_{\bar w} b = 0.
\label{eq:sol_qlump}
\eeq
Thus the solution is given by
\beq
b(t,x^1,x^2) = e^{\pm i m t}\, \tilde b(w),
\eeq
where $\tilde b(w)$ can be any holomorphic function of $w$ 
as in the case of the static configuration Eq.~(\ref{eq:holo_map}).
Note that the Q-lump solution with the minimal winding number 
($\tilde b = \frac{\lambda}{w}$) has infinite energy because 
its Noether charge density has an asymptotic tail $\sim 1/|\lambda^2w|$ ~\cite{Abraham:1991ki}. The minimal configuration with finite energy,
for example, is given by $\tilde b = \frac{\lambda}{w-w_1} - \frac{\lambda}{w - w_2}$.

Let us next see the effects from the higher derivative corrections: 
Do the Q-lumps receive corrections from the higher order terms?
Indeed, as one can easily check, the solution given in Eq.~(\ref{eq:sol_qlump}) solves the full equations of motion including the fourth order corrections.
In the presence of the higher order corrections, 
the energy density can be rewritten as follows 
\beq
E^{(4)} &=& \frac{4\pi c}{g^2}\int d^2x\ \frac{2\left(|\dot b|^4 - |(\p_i b)^2-m^2 b^2|^2\right) + |\dot b^2-(\p_ib)^2 + m^2 b^2|^2}{(1+|b|^2)^4}
\nonumber\\
&=&  \frac{32\pi c}{g^2}\int d^2x\ 
\frac{m^2\left(\bar{\tilde b}^2 \p_w \tilde b \p_{\bar w}\tilde b + \tilde b^2 \p_w \bar{\tilde b} \p_{\bar w} \bar{\tilde b} \right) 
- 2 | \p_w \tilde b \p_{\bar w} \tilde b |^2}{(1+|\tilde b|^2)^4}.
\eeq 
Since $\tilde b$ is a holomorphic function in $w$, 
we can immediately conclude that $E^{(4)} = 0$ 
for the Q-lump solutions $b = e^{imt}\tilde b(w)$.
Therefore, the BPS Q-lump solutions and 
their masses are not modified by the higher order corrections.

%%%%%%%%%%%%%%%%%%%%%%%%%%%%%%%%%%%%%%%%%%%
%\newpage
\section{Summary and Discussion}\label{sec:sum}

We have proposed a systematic method to obtain 
higher derivative terms in the low-energy effective theories on solitons.
We have applied our method to a single non-Abelian vortex 
and have obtained four-derivative terms 
in the ${\mathbb C}\times {\mathbb C}P^{N-1}$ model 
on the vortex world-volume. 
We have compared our four-derivative terms 
with the Nambu-Goto action 
and the Faddeev-Skyrme model. 
The action for the translational moduli $Z$ coincides with 
the Nambu-Goto action and the terms for the orientational moduli $b^i$ 
coincide with those for the supersymmetric extension 
of the Faddeev-Skyrme term.
We have also shown that the contribution 
from the four-derivative terms disappears for 1/4 BPS states of 
instantons trapped inside a non-Abelian vortex 
and consequently the solutions are not modified 
in the presence of the four derivative terms.

In this paper, we have derived the four derivative terms. 
In principle we can go on to any order in our formalism.
The sixth order is considered to be prominently important.
We have confirmed that the effective action of the translational zero modes $Z$ at the fourth order is consistent with the Nambu-Goto action. 
On the other hand, in the width expansion from the Nambu-Goto action, 
the first correction term written as the extrinsic curvature squared 
starts from the sixth derivatives in the derivative expansion 
\cite{Polyakov:1986cs}. 
In field theory calculation, 
it seems that there is no agreement on the signature of that term 
\cite{higher}.  
Therefore, we can in principle determine that term for the BPS case.

In this paper we have studied {\it local} non-Abelian vortices 
which exist in the theory with the number $N_{\rm F}$ of flavors 
equals to the number $N_{\rm C}$ of color.
When the theory has more flavors, $N_{\rm F}>N_{\rm C}$, 
vortices are called {\it semi-local} \cite{Vachaspati:1991dz}. 
It is known in this case that the orientational zero modes 
of a single vortex is non-normalizable, {\it i.e.}, 
the integration over the codimensions diverges \cite{Shifman:2006kd},  
unless the size modulus is zero and the vortex shrinks to a local vortex 
\cite{Eto:2007yv}. 
For two vortices, the relative orientational zero modes 
are normalizable even with a non-zero size moduli, 
while the overall orientational zero modes are 
non-normalizable \cite{Eto:2007yv}. 
Although we can formally extend our method to semi-local vortices,  
we should check if there exists a divergence in four derivative terms 
even for normalizable moduli. 
Remember that in the derivative expansion, derivatives are assumed
to be less than the lowest mass $m$ of the mass spectrum in the vacuum. 
The existence of the vacuum moduli 
for $N_{\rm F}>N_{\rm C}$ implies that $m=0$ and 
hence the convergence radius of the derivative expansion 
seems to be zero. 
Therefore, it is interesting to see if 
there exists a special mechanism which justifies 
the derivative expansion for semi-local vortices.
Non-Abelian vortices were extended to arbitrary gauge groups $G$  
in the form of ${(U(1) \times G)/ C(G)}$ with the center $C(G)$ of $G$
\cite{Eto:2008yi}. 
Especially the cases of $G=SO(N),\,USp(2N)$ have been 
studied in detail \cite{Eto:2008qw,Ferretti:2007rp}. 
We can straightforwardly extend our analysis 
to the cases of arbitrary gauge groups
but we should be careful to the normalizability 
since they are semi-local vortices in general. 

Although we have studied BPS vortices in supersymmetric gauge theories, 
our method to obtain higher derivative corrections by solving equations of motion for massive fields is robust and can be extended to non-supersymmetric theories.
For instance, we can apply it to non-Abelian vortices in 
non-supersymmetric theory \cite{Gorsky:2004ad}. 
In reality, non-Abelian vortices exist in high-density QCD 
which may be realized in the core of neutron stars \cite{Balachandran:2005ev}. 
In this case, the low-energy world-sheet theory on the vortex is described by the bosonic ${\mathbb C}P^2$ model at the leading order \cite{Nakano:2007dr}. 
The four-derivative correction to it should be important 
especially for the fate of confined monopoles which have been recently 
shown to exist as kinks on the vortex \cite{Gorsky:2011hd}.
This is because they are non-BPS and higher derivative terms do not vanish 
automatically, 
unlike BPS instantons discussed in Sec.\,\ref{sec:inst}.

Finally our method is general so that we can apply it to other BPS solitons such as domain walls, monopoles and instantons, 
or non-BPS solitons such as Skyrmions. 
In the same spirit, 
four derivative term in the form of the Skyrme term was obtained in the effective theory of non-Abelian domain walls 
\cite{Eto:2005cc}, and
four derivative terms for collective coordinates of a rotating Skyrmion 
were calculated \cite{Hata:2010vj}.

%%%%%%%%%%%%%%%%%%%%%%%%%%%%%%%%%%%%%%%%%%%%%%%%%%%%%%%%%%%%%%%%%%%%%%
\section*{Acknowledgments}

We would like to thank Kaneyasu Asakuma for a discussion 
in the early stage of this work. 

The work of M.~E.~, M.~N.~and N.S.~are supported in part by
Grant-in Aid for Scientific Research 
No. 23740226 (M.E.), No. 23740226 (M.N.), 
the ``Topological Quantum Phenomena''Grant-in Aid for Scientific Research
on Innovative Areas (No. 23103515) (M.N.), 
No.~21540279 (N.S.) and No.~21244036 (N.S.) from the Ministry of Education, 
Culture, Sports, Science and Technology(MEXT) of Japan,
and by Japan Society for the Promotion of Science (JSPS)
and Academy of Sciences of the Czech Republic (ASCR) under
the Japan - Czech Republic Research Cooperative Program
(M.E.~and N.S.).

%%%%%%%%%%%%%%%%%%%%%%%%%%%%%%%%%%%%%
\begin{appendix}
\section{O(3) model}
From the isomorphism 
${\mathbb C}P^1 \simeq S^2 \simeq O(3)/O(2)$, 
the ${\mathbb C}P^1$ model is equivalent to the $O(3)$ model.
In order to see this equivalence, let us introduce a three vector
${\bf n} = (n_1,n_2,n_3)$ by
\beq
{\bf n} &=& \ba{cc} \frac{1}{\sqrt{1+|\beta|^2}}, & \frac{\bar\beta}{\sqrt{1+|\beta|^2}} \ea \vec{\sigma} 
\ba{c} \frac{1}{\sqrt{1+|\beta|^2}} \\ \frac{\beta}{\sqrt{1+|\beta|^2}} \ea \nonumber \\
&=& \left( {\beta + \bar\beta \over 1+|\beta|^2}, 
 -i{\beta - \bar \beta \over 1+|\beta|^2}, 
 {1 -|\beta|^2 \over 1+|\beta|^2}
   \right) \label{eq:n}
\eeq
which satisfies the constraint 
\beq
{\bf n}^2 = 1.
\eeq
Conversely, $\beta$ is the stereographic coordinate, given by 
\beq
 \beta = {n_1 + i n_2 \over 1 + n_3}
 = {1 - n_3 \over n_1 - i n_2}.
\eeq
The kinetic term becomes 
\beq 
 {\partial_\mu \beta \partial^{\mu} \bar\beta \over (1+|\beta|^2)^2}
 = {1\over 2} \partial_{\mu} {\bf n} \cdot \partial^{\mu} {\bf n} ,
\eeq
and the field strength can be rewritten as
\beq
 f_{\mu\nu} 
 = {\bf n} \cdot (\partial_{\mu} {\bf n} \times \partial_{\nu}{\bf n}).
\eeq
Therefore the Skyrme-Faddeev term and the other four derivative term become
\beq 
f_{\mu\nu} f^{\mu \nu} ~~
&=& (\partial_{\mu} {\bf n} \times \partial_{\nu} {\bf n})^2 , \phantom{{(\partial_{\mu} \beta \partial^{\mu} \bar\beta)^2 \over (1+|\beta|^2)^4}} \\
{(\partial_{\mu} \beta \partial^{\mu} \bar\beta)^2 \over (1+|\beta|^2)^4}
&=& {1\over 4} (\partial_{\mu} {\bf n} \cdot \partial^{\mu} {\bf n})^2 ,
\eeq
respectively. 
The total four derivative terms in 
Eq.\,(\ref{eq:four-deriv-general})
can be rewritten as
\beq
 {\cal L}_4 
 = c_1 (\partial_{\mu} {\bf n} \times \partial_{\nu} {\bf n})^2 
 + {c_2 \over 4} (\partial_{\mu} {\bf n} \cdot \partial^{\mu} {\bf n})^2.
\eeq

%%%%%%%%%%%%%%%%%%%%%%%%%%%%%%%%%%%%%%%%%%%%%%%%%%%%%%%%%%%%%%
\section{Gauged linear sigma model and higher derivative corrections}\label{appendix:higher_CP}
In this section, we consider a gauged linear sigma model 
with a higher derivative term
which reproduces the orientational part of 
the higher derivative terms in Eq.\,\eqref{eq:sum}. 
The model is described by $N$ charged scalar fields $h=(h_1,\cdots,h_N)$ 
coupled to a $U(1)$ auxiliary gauge field $a_\alpha$. 
The Lagrangian is given by
\beq
\mathcal L ~=~ \D_\alpha h (\D^\alpha h)^\dagger + d \, \D_\alpha h (\D_\beta h)^\dagger \, \D^\alpha h (\D^\beta h)^\dagger + \lambda \left( h h^\dagger - v^2 \right),
\label{eq:abelian_L}
\eeq
where $d$ is a constant and $\lambda$ is a Lagrange multiplier 
for the constraint $h h^\dagger = v^2$, which is solved by
\beq
h = \frac{v}{\sqrt{1+|b_i|^2}} (b_1,\cdots,b_{N-1},1).
\eeq
The covariant derivative is defined by 
$\D_\alpha h = (\p_\alpha + i a_\alpha) h$. 
The equation of motion for the auxiliary gauge field 
$a_\alpha$ can be solved as
\beq
a_\alpha = -\frac{i}{2} \frac{b_i \p_\alpha \bar b_i - \p_\alpha b_i \bar b_i}{1 + |b_i|^2}.
\label{eq:a_alpha}
\eeq
This solution is independent of the parameter $d$. 
In other words, the higher derivative term in
Eq.\,\eqref{eq:abelian_L} does not change the solution 
for the auxiliary gauge field. 
We can easily show that Eq.\,\eqref{eq:a_alpha} is the solution 
by using the following relations
\beq
\D_\alpha h h^\dagger = h (\D_\alpha h)^\dagger = 0.
\eeq
Substituting Eq.\,\eqref{eq:a_alpha} back into 
the Lagrangian Eq.\,\eqref{eq:abelian_L}, we obtain
\beq
\mathcal L = v^2 g_{i \bar j}^{\rm FS} \p_\alpha b^i \p^\alpha \bar b^j + v^4 d (g_{i \bar j}^{\rm FS} \p_\alpha b^i \p_\beta \bar b^j) (g_{k \bar l}^{\rm FS} \p^\alpha b^k \p^\beta \bar b^l).
\eeq
Therefore, if $v^2 = \frac{4\pi}{g^2}$ and $d= \frac{g^2}{4\pi} c$, 
this Lagrangian coincides with the orientational part of 
the vortex effective action Eq.\,\eqref{eq:sum}. 

\section{Regularity of the second order solution}\label{sec:singularity}
In this section, we show that there is no singularity in 
the first component of the solution $\Delta \Phi^{(2)}$ 
given in Eq.\,\eqref{eq:sol_Phi2}.  
Here we assume that $Z_I \not = Z_J \, ( I \not = J ) $, 
so that the constant matrix 
$\Psi^I \equiv \left[ (z-Z_I) H_0^{-1} \right]_{z=Z_I}$ is well-defined. 
Then the singular behavior around the 
$I$-th vortex position $z=Z_I$ can be written as
\beq
\frac{1}{\bar z - \bar Z_I} \left[ \Psi_I \bar \nabla_i \frac{\p}{\p \bar \phi^j} \left(\Omega \p_z \Omega^{-1} \right) \right]_{z=Z_I}^\dagger + \mathcal O(1).\label{eq:singular}
\eeq
We can show that
\beq
\left[ \bar \nabla_i \frac{\p}{\p \bar \phi^j} \left(\Omega \p_z \Omega^{-1} \right) \right]_{z=Z_I} &=& \bar \nabla_i \frac{\p}{\p \bar \phi^j} \left[ \, \Omega \p_z \Omega^{-1} \right]_{z=Z_I} - \bar \nabla_i \left[ \frac{\p Z_I}{\p \bar \phi^j} \, \p_{\bar z} (\Omega \p_z \Omega^{-1} ) \right]_{z=Z_I} \\
&-& \frac{\p \bar Z_I}{\p \bar \phi^i} \frac{\p}{\p \bar \phi^j} \left[ \p_{\bar z} ( \Omega \p_z \Omega^{-1} ) \right]_{z=Z_I} +  \frac{\p \bar Z_I}{\p \bar \phi^i} \frac{\p Z_I}{\p \bar \phi^j} \left[ \p_{\bar z}^2 (\Omega \p_z \Omega^{-1}) \right]_{z=Z_I}, \notag 
\eeq
where we have used the following identity for any function of the form
$f(z, \bar z, \phi^i,\bar \phi^i)$
\beq
\left[ \frac{\p}{\p \bar \phi^i} f \right]_{z=Z_I} = \frac{\p}{\p \bar \phi^i} \left[ f \right]_{z=Z_I} - \frac{\p \bar Z_I}{\p \bar \phi^i} \left[ \p_{\bar z} f \right]_{z=Z_I}.
\eeq
Then, the singular part can be rewritten into the following form
\begin{eqnarray}
- \frac{g^2v^2}{4} \frac{1}{\bar z - \bar Z_I} \left[ \Psi_I \bar \nabla_i 
\frac{\p }{\p \bar \phi^j} \left( B_I - \bar Z_I {\bf 1}_N \right) \right]^\dagger,
\quad {\rm with} \quad 
B_I \equiv -\frac{4}{g^2v^2} \left[ \Omega \p_z \Omega^{-1} \right]_{z=Z_I},
\end{eqnarray}
where we used the following relations 
which can be derived from the master equation Eq.\,\eqref{eq:master}:
\begin{eqnarray}
\Psi_I\p_{\bar z} \left[ \Omega \p_z \Omega^{-1} \right]_{z=Z_I} =
-\frac{g^2v^2}{4} \Psi_I, \qquad 
\Psi_I\p_{\bar z}^{2} \left[ \Omega \p_z \Omega^{-1}\right]_{z=Z_I}=0.
\end{eqnarray}
Now let us use the explicit form of 
the generic moduli matrix Eq.\eqref{eq:genericH0}. 
Then, $\Psi_I$ is given by
\beq
\Psi_I ~=~ \prod_{I \not = J} \frac{1}{Z_I - Z_J} \ba{cc} \ 1 & \ 0 \\ -\vec b & \ \mathbf 0 \ea. 
\eeq
By assuming that the matrix $B_I$ takes the form
\beq
B_I \equiv \ba{cc} p_I & (\vec q_I)^{\rm T} \\ \vec r_I & s_I \ea, 
\eeq
we can rewrite the singular part as
\beq
- \frac{g^2 v^2}{4} \frac{1}{\bar z - \bar Z_I} \left[ \prod_{J \not = I} \frac{1}{Z_I - Z_J} \ba{cc} \ 1 & \ 0 \\ -\vec b & \ \mathbf 0 \ea \bar \nabla_i \frac{\p}{\p \bar \phi^j} \ba{cc} p_I - \bar Z_I & (\vec q_I)^{\rm T} \\ \vec r_I & s_I - Z_I \ea \right]^\dagger.
\eeq
We can show that $\bar \nabla_i \frac{\p}{\p \bar \phi^j} (\bar Z_I - p_I) = \bar \nabla_i \frac{\p}{\p \bar \phi^j} \vec q_I = 0$ as follows.
It has been shown that the matrix $B_I$ is related to 
the moduli space metric as \cite{Fujimori:2010fk}
\begin{eqnarray}
 g_{i\bar j} &=& \pi v^2 \sum_{I=1}^{k}
\left( \frac{\p Z_I}{\p \phi^i} \frac{\p \bar Z_I}{\p \bar \phi^j}
+{\rm Tr} \left[ \frac{\p H_0}{\p \phi^i} \Psi_I \frac{\p B^I}{\p \bar \phi^j}\right]_{z=Z_I} \right) \label{eq:metricformula} \\
&=& \pi v^2 \sum_{I=1}^{k} \left( \frac{\p Z_I}{\p \phi^i} \frac{\p}{\p \bar \phi^j} (\bar Z_I - p_I) + \prod_{J \not = I} \frac{1}{Z_I - Z_J} \left[ \frac{\p \vec b_I}{\p \phi^i} - \sum_{J=1}^k \frac{\p Z_J}{\p \phi^i} D_J{}^I \vec b_J  \right] \cdot \frac{\p \vec q_I}{\p \bar \phi^j} \right), \notag
\end{eqnarray}
where $D_J{}^I \equiv \left[ \p_z e_J(z) \right]_{z=Z_I}$.
Since the vortex moduli space is a K\"ahler manifold, 
there exist a K\"ahler potential $\mathcal K$ such that
\beq
g_{i \bar j} ~=~ \frac{\p \mathcal K}{\p \phi^i \p \bar \phi^j}. 
\eeq
By comparing this equation with Eq.\eqref{eq:metricformula}, 
we can read $p_I$ and $\vec q_I$ as  
\begin{eqnarray}
\frac{\p}{\p \bar \phi^j} (\bar Z_I - p_I) &=& \ \ \frac{1}{\pi v^2} \left[ \frac{\p {\cal K}}{\p \bar \phi^j \p Z_I} - \sum_{J=1}^k D_I{}^J \vec b_I \cdot \frac{\p {\cal K}}{\p \bar \phi^j \p \vec b_J} \right], \\ 
\frac{\p \vec q_I}{\p \bar \phi^j} \hs{7} &=& \ \ \frac{1}{\pi v^2} \prod_{J \not = I} (Z_I - Z_J) \frac{\p {\cal K}}{\p \bar \phi^j \p \vec b_I},
\end{eqnarray}
Since the moduli space metric is covariantly constant, it follows that 
\begin{eqnarray}
\bar \nabla_i \frac{\p}{\p \bar \phi^j} (\bar Z_I - p_I) ~=~ \bar \nabla_i \frac{\p \vec q_I}{\p \bar \phi^j} ~=~ 0.
\end{eqnarray}
This shows that there is no singularity in the solution Eq.\,\eqref{eq:sol_Phi2}. 
\end{appendix}

\newpage
%%%%%%%%%%%%%%%%%%%%%%%%%%%%%%%%%%%%%%%%%%%%%%%%%%%%%%%%%%%%%%

\end{document}